\documentclass[aps,prd,twocolumn,superscriptaddress,nofootinbib,showpacs,letterpaper,eqsecnum]{revtex4}

\usepackage{mathptmx}
\usepackage{showlabels}
\usepackage{verbatim}
\usepackage[usenames,dvipsnames]{color}
\usepackage[pdftex]{graphicx}
\usepackage[english]{babel}
\usepackage{amsmath,amssymb}
\usepackage[colorlinks=true,citecolor=blue]{hyperref}

\begin{document}

\title{Quasinormal-mode spectrum of Kerr black holes and its geometric 
interpretation}
\author{Huan Yang}
\affiliation{Theoretical Astrophysics 350-17, California Institute of Technology, Pasadena, CA 91125, USA}
\author{David A.\ Nichols}
\affiliation{Theoretical Astrophysics 350-17, California Institute of Technology, Pasadena, CA 91125, USA}
\author{Fan Zhang}
\affiliation{Theoretical Astrophysics 350-17, California Institute of Technology, Pasadena, CA 91125, USA}
\author{Aaron Zimmerman}
\affiliation{Theoretical Astrophysics 350-17, California Institute of Technology, Pasadena, CA 91125, USA}
\author{Zhongyang Zhang}
\affiliation{Department of Physics and Astronomy, The University of Mississippi, University, MS 38677, USA}
\author{Yanbei Chen}
\affiliation{Theoretical Astrophysics 350-17, California Institute of Technology, Pasadena, CA 91125, USA}

\begin{abstract}
There is a well-known, intuitive geometric correspondence between 
high-frequency quasinormal modes of Schwarzschild black holes and null 
geodesics that reside on the light-ring (often called spherical photon 
orbits): the real part of the mode's frequency relates to the geodesic's
orbital frequency, and the imaginary part of the frequency corresponds to 
the Lyapunov exponent of the orbit.
For slowly rotating black holes, the quasinormal-mode's real frequency is a 
linear combination of a the orbit's precessional and orbital frequencies, but 
the correspondence is otherwise unchanged.
In this paper, we find a relationship between the quasinormal-mode frequencies 
of Kerr black holes of arbitrary (astrophysical) spins and general spherical 
photon orbits, which is analogous to the relationship for slowly rotating holes.
To derive this result, we first use the WKB approximation to compute accurate 
algebraic expressions for large-$l$ quasinormal-mode frequencies.
Comparing our WKB calculation to the leading-order, geometric-optics 
approximation to scalar-wave propagation in the Kerr spacetime, we then draw a 
correspondence between the real parts of the parameters of a quasinormal mode
and the conserved quantities of spherical photon orbits.
At next-to-leading order in this comparison, we relate the imaginary parts of
the quasinormal-mode parameters to coefficients that modify the amplitude of
the scalar wave.
With this correspondence, we find a geometric interpretation to two features of
the quasinormal-mode spectrum of Kerr black holes:
First, for Kerr holes rotating near the maximal rate, a large number of modes
have nearly zero damping; we connect this characteristic to the fact that a 
large number of spherical photon orbits approach the horizon in this limit.
Second, for black holes of any spins, the frequencies of specific sets of modes
are degenerate; we find that this feature arises when the spherical photon 
orbits corresponding to these modes form closed (as opposed to ergodically 
winding) curves. 
\end{abstract}


\maketitle

\section{Introduction}

Quasinormal modes (QNMs) of black-hole spacetimes are the characteristic modes 
of linear perturbations of black holes that satisfy an outgoing boundary 
condition at infinity and an ingoing boundary condition at the horizon (they 
are the natural, resonant modes of black-hole perturbations). 
These oscillatory and decaying modes are represented by complex characteristic
frequencies $\omega = \omega_R - i \omega_I$, which are typically indexed by 
three numbers, $n$, $l$, and $m$. 
The overtone number $n$ is proportional to the decay rate of the perturbation, 
and $l$ and $m$ are multipolar indexes of the angular eigenfunctions of the 
QNM.

\subsection{Overview of quasinormal modes and their geometric interpretation}

Since their discovery, numerically, in the scattering of gravitational waves in
the Schwarzschild spacetime by Vishveshwara \cite{Vishveshwara1970}, QNMs have 
been thoroughly studied in a wide range of spacetimes, and they have found many
applications.
There are several reviews \cite{KokkotasSchmidt, Nollert, Ferrari2008, berti, 
Konoplya2011} that summarize the many discoveries about QNMs.
They describe how QNMs are defined, the many methods used to calculate
QNMs (e.g., estimating them from time-domain solutions \cite{Davis1971}, 
using shooting methods in frequency-domain calculations 
\cite{ChandraDetweiler1975}, approximating them with inverse-potential
approaches \cite{Ferrari1984} and WKB methods \cite{schutz, Iyer2}, and 
numerically solving for them with continued-fraction techniques 
\cite{Leaver, Nollert1993}), and the ways to quantify the excitation of QNMs 
(see, e.g., \cite{Leaver1986, sun}).  
They also discuss the prospects for detecting them in gravitational waves using
interferometric gravitational-wave detectors, such as LIGO \cite{LIGO} and 
VIRGO \cite{VIRGO}, and for inferring astrophysical information from them (see,
e.g., \cite{echeverria, Flanagan1998} for finding the mass and spin of black 
holes using QNMs, \cite{Buonanno2007, Berti2007} for quantifying the excitation
of QNMs in numerical-relativity simulations binary-black-hole mergers,  and 
\cite{Dreyer2004, Berti2006} for testing the no-hair theorem with QNMs).
There have also been several other recent applications of QNMs. 
For example, Zimmerman and Chen \cite{aaron} (based on work by Mino and Brink 
\cite{jeandrew}) study extensions to the usual spectrum of modes generated in 
generic ringdowns.
Dolan and Ottewill use eikonal methods to approximate the modal wave function, 
and they use these functions to study the Green's function and to help 
understand wave propagation in the Schwarzschild spacetime 
\cite{Sam3,Sam,Sam2}.

Although QNMs are well-understood and can be calculated quite precisely, it 
remains useful to develop intuitive and analytical descriptions of these modes.
An important calculation in this direction was performed by Ferrari and 
Mashhoon \cite{Ferrari1984}, who showed that for a Schwarzschild black hole, 
the QNM frequency (which depends only on a multipolar index $l$ and an overtone
index $n$) can be written in the eikonal (or geometric-optics) limit, 
$l \gg 1$, as 
\begin{equation}
\label{MashoonFreq}
\omega \approx (l+1/2)\Omega - i \gamma_L (n+1/2) \, .
\end{equation}
The quantities $\Omega$ and $\gamma_L$ are, respectively, the Keplerian 
frequency of the circular photon orbit and the Lyapunov exponent of the orbit, 
the latter of which characterizes how quickly a congruence of null geodesics on
the circular photon orbit increases its cross section under infinitesimal 
radial perturbations \cite{Cardoso2009,Sam2}. 
Equation (\ref{MashoonFreq}) hints at an intriguing physical description of 
QNMs: for modes with wavelengths much shorter than the background curvature, 
the mode behaves as if it were sourced by a perturbation that orbits on and 
diffuses away from the light ring on the time scale of the Lyapunov exponent. 

Ferrari and Mashhoon \cite{Ferrari1984} also derived an analogous result to 
Eq.~\eqref{MashoonFreq} for slowly rotating black holes.
They showed for $l\gtrsim m \gg 1$, the real part of the frequency is given by
\begin{equation}
\Omega \approx \omega_{\rm orb} + \frac{m}{l + 1/2} \omega_{\rm prec} \, ,
\end{equation}
where $\omega_{\rm orb}$ is now the Keplerian orbital frequency for the 
circular photon orbit and $\omega_{\rm prec}$ is the Lense-Thiring-precession 
frequency of the orbit (which arises because of the slow rotation of the black 
hole). 
The term proportional to $\omega_{\rm prec}$ also has a simple geometric-optics
interpretation.
Inertial frames near the high-frequency wave at the light ring are dragged 
with respect to inertial frames at infinity, and this frame dragging causes
the perturbation's orbit to precess about the spin axis of the black hole with 
a frequency, $\omega_{\rm prec}$. 
If the orbit is inclined at an angle of $\sin^2 \theta = m^2/l(l+1)$ (the ratio
of angular momenta $L_z^2/L^2$ for quantized waves in flat space), then the 
projection of the precessional velocity onto the orbital plane scales the 
precessional frequency by a factor of $\sim m/(l+ 1/2)$.

Why the QNM frequency is multiplied by $(l+1/2)$ is a feature that we will 
explain in greater detail in this paper.
Intuitively, this term arises because the in the high-frequency limit, any 
wavefront traveling on null orbits will have an integral number of oscillations
in the $\theta$ and $\phi$ directions. 
For the wave to be periodic and single-valued, there must be $m$ oscillations 
in the $\phi$ direction.
For the $\theta$ direction, it is a Bohr-Sommerfeld quantization condition that
requires $l - |m| +1/2$ oscillations in this direction, which implies that 
there should be a net spatial frequency of roughly $(l+1/2)$. 
This increases the frequency of the radiation seen far from the hole by the 
same factor.

From this intuitive argument, we expect that the real part of the mode should
be
\begin{equation}
\label{FreqSum}
\omega_{R} = L \left(\omega_{\rm orb}+ \frac{m}{L}\omega_{\rm prec} \right)
\, ,
\end{equation}
where we define $L = l + 1/2$. 
In this paper, we will show that an equation of the form of Eq.\ 
(\ref{FreqSum}) does, in fact, describe the QNM frequencies of Kerr black holes
of arbitrary astrophysical spins (and it recovers the result of Ferrari and 
Mashhoon for slowly spinning black holes).
As we mention in the next part of this section, the exact details of the 
correspondence between QNMs and photon orbits is richer for rapidly rotating
black holes than for slowly rotating or static black holes.

\subsection{Methods and results of this article}

\begin{figure*}
\includegraphics[width=8.7in, bb=0 0 1800 276, clip]{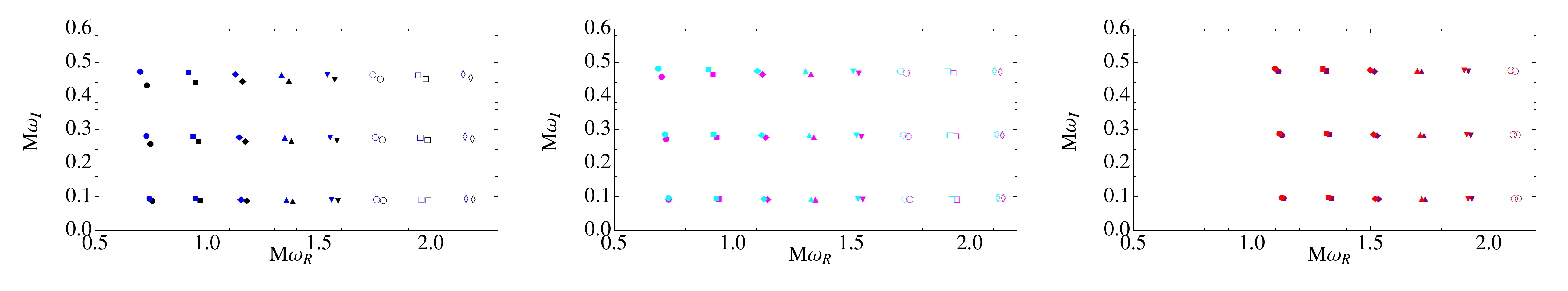}
\caption{Low-overtone QNM spectrum of three Kerr black holes of different spins
with approximate degeneracies in their spectra.  
From left to right, we plot the three lowest-overtone QNM excitations for (i)
$a/M=0.69$ in which $(l,m) = (j,2)$ are black symbols and $(l',m')=(j+1,-2)$ 
are blue symbols, where $j=3,\ldots,9$; (ii) $a/M=0.47$ in which $(l,m)=(j,3)$ 
are magenta symbols and $(l',m')=(j+1,-3)$ are cyan symbols, where 
$j=3,\ldots,9$; (iii) $a/M=0.35$ in which $(l,m) = (j,4)$ are red symbols and 
$(l',m')=(j+1,-4)$ are purple symbols, where $j=5,\ldots,10$.  
For these spin parameters, the mode with positive values of $m$ and 
$\omega_{R}$ (a corotating mode) of index $l$ is approximately degenerate with 
the mode with $m'=-m,$ and $\omega_{R}$ (a counterrotating mode) of index 
$l'=l+1$.}
\label{numdeg}
\end{figure*}

To derive Eq.\ \eqref{FreqSum} requires that we develop a geometric-optics 
interpretation of the QNMs of Kerr black holes with arbitrary astrophysical 
spins.
Finding the correspondence requires two steps: first, we need to calculate the 
approximate frequencies using the WKB method; next, we must articulate a 
connection between the mathematics of waves propagating in the Kerr spacetime 
in the geometric-optics approximation and those of the WKB approximation (the 
first step).
Finally, with the geometric-optics description of QNMs, we can make a physical
interpretation of the spectrum (for example, the degeneracy or the lack of 
damping in the extremal limit).

In Sec.\ \ref{sec:WKB}, we describe how we solve the eigenvalue problem that 
arises from separating the Teukolsky equation \cite{Teukolsky} (a linear 
partial differential equation that describes the evolution of scalar, vector, 
and gravitational perturbations of Kerr black holes) into two nontrivial linear
ordinary differential equations.
The two differential equations, the radial and angular Teukolsky equations, 
share two unknown constants---the frequency, $\omega$, and the angular 
separation constant, $A_{lm}$---that are fixed by the boundary conditions that
the ordinary differential equations must satisfy (ingoing at the horizon and 
outgoing at infinity for the radial equation, and well-behaved at the poles
for the angular equation).
The goal of the WKB method is to compute the frequency and separation constant
approximately.

Although there has been work by Kokkotas \cite{Kokkotas} and Iyer and Seidel 
\cite{Iyer} using WKB methods to compute QNM frequencies of rotating black 
holes, their results were limited to slowly rotating black holes, because they 
performed an expansion of the angular separation constant, $A_{lm}$, for small,
dimensionless spin parameters, $a/M$, and only applied the WKB method to the 
radial Teukolsky equation to solve for the frequency. 
In a different approach, Dolan developed a matched-expansion formalism for Kerr
black holes of arbitrary spins that can be applied to compute the frequency of
QNMs, but only for modes with $l=|m|$ and $m=0$ \cite{Sam}. 

Therefore, it remains an outstanding problem to compute a WKB approximation 
to the quasinormal modes of Kerr black holes of large spins and for any 
multipolar index $m$.
In Sec.\ \ref{sec:WKB}, we solve the joint eigenvalue problem of the radial and
angular Teukolsky equations by applying a change of variables to the angular 
equation that brings it into the form of a bound-state problem in quantum 
mechanics.
Applying the WKB method to the angular equation, we arrive at a Bohr-Sommerfeld
condition that constrains the angular constant in terms of the frequency (and
the indices $l$ and $m$). 
Simultaneously, we can analyze the radial equation in the WKB approximation, 
and the two equations together define an system of integral equations, which 
can be solved for the eigenvalues. 
When we expand the Bohr-Sommerfeld condition in a Taylor series in terms of the
numerically small parameter, $a\omega/l$, the system of integral equations 
reduces to an algebraic system (which, in turn, leads to a simpler expression
for the frequency).
The approximate frequency agrees very well with the result that includes all 
powers of $a\omega/l$, and, in the eikonal limit, it is accurate to order 
$1/l$ for Kerr black holes of arbitrary spins, for modes with any value of $m$,
and for both the real and the imaginary parts of the frequency.

To interpret the WKB calculation of Sec.\ \ref{sec:WKB} in the language of 
propagating waves in the geometric-optics limit within the Kerr spacetime, we
analyze waves around a Kerr black hole in Sec.\ \ref{sec:GeomOptics} using the 
geometric-optics approximation and the Hamilton-Jacobi formalism.
We confirm that the leading-order pieces of the WKB frequencies and angular 
constants correspond to the conserved quantities of the leading-order, geodesic 
behavior of the geometric-optics approximation (specifically, the real part of 
$\omega$, the index $m$, and the real part of $A_{lm}$ are equivalent to the 
energy $\mathcal E$, the z-component of the specific angular momentum $L_z$, 
and Carter's constant $\mathcal Q$ plus $L_z^2$, respectively).
The specific geodesics corresponding to a QNM are, in fact, spherical photon 
orbits.
The next-to-leading-order WKB quantities (the imaginary parts of $\omega$ and
$A_{lm}$) correspond to dispersive, wavelike corrections to the geodesic motion
(they are the Lyapunov exponent and the product of this exponent with the 
change in Carter's constant with respect to the energy). 
Table \ref{tb:GeoMatch} in Sec.\ \ref{sec:GeomOptics} summarizes this 
geometric-optics correspondence.

In Sec.\ \ref{sec:Spectra}, we make several observations about features of the
QNM spectrum of Kerr black holes that have simple geometric interpretations.
First, we find that for extremal Kerr black holes, a significant fraction of
the QNMs have a real frequency proportional to the angular frequency of the 
horizon and a decay rate that rapidly falls to zero; we explain this
in terms of a large number of spherical photon orbits that collect on the 
horizon for extremal Kerr holes. 
Second, we expand the WKB expression for the real part of the frequency as in
Eq.~\eqref{FreqSum}, and we interpret these terms as an orbital and a 
precessional frequency of the corresponding spherical photon orbit.
These two frequencies depend on the spin of the black hole and the value of
$m/L$ very weakly for slowly-rotating black holes, though quite strongly when
the spin of the black hole is nearly extremal.
Finally, we use the geometric-optics interpretation given by Eq.\
\eqref{FreqSum} to explain a degeneracy in the QNM spectrum of Kerr black 
holes, in the eikonal limit, which also manifests itself, approximately, for 
small $l$ (see Fig.\ \ref{numdeg}).
The degeneracy occurs when the orbital and precession frequencies, 
$\omega_{\rm orb}$ and $\omega_{\rm prec}$ are rationally related (i.e.,
$\omega_{\rm orb}/\omega_{\rm prec} = p/q$ for integers $p$ and $q$) for a 
hole of a specific spin parameter, and when the corresponding spherical 
photon orbits close.
By substituting this result into Eq.\ (\ref{FreqSum}) one can easily see
that modes with multipolar indexes $l$ and $m$ become degenerate with those of
indexes $l' = l+k q$ and $m' = m - k p$ for any non-negative integer $k$, in
the eikonal limit (note that in Fig.\ \ref{numdeg}, we show an approximate 
degeneracy for $k=1$ and for three spin parameters, such that $q/p=1/4$, $1/6$,
and $1/8$, respectively.)

To conclude this introduction, we briefly summarize the organization of this
paper:
In Sec.~\ref{sec:WKB}, we review the Teukolsky equations, and we then describe
the WKB formalism that we use to calculate an accurate approximation to the 
angular eigenvalues $A_{lm}=A_{lm}^R+iA_{lm}^I$ and QNM frequencies 
$\omega = \omega_R - i \omega_I$, in the eikonal limit $L\gg 1$ and for holes
of arbitrary spins. 
We verify the accuracy of our expressions in Sec.~\ref{sec:WKBaccuracy} by 
comparing the WKB frequencies to exact, numerically calculated frequencies.
In Sec.~\ref{sec:GeomOptics}, we develop a correspondence between the WKB 
calculation and mathematics of wave propagation within the Kerr spacetime in 
the geometric-optics limit, using the geometric-optics approximation and the
Hamilton-Jacobi formalism.
At leading-order, the QNM frequencies and angular eigenvalues correspond to the
conserved quantities of motion in the Kerr spacetime for spherical photon 
orbits; at next-to-leading order in the geometric-optics approximation, we 
connect the the decaying behavior of the QNMs to dispersive behaviors of the 
waves. 
Finally, in Sec.~\ref{sec:Spectra}, we interpret aspects of the QNM spectrum 
geometrically, such as the vanishing of the damping rate for many modes of 
extremal black holes, the decomposition of the frequency into orbital and 
precessional parts, and the degeneracies in the QNM frequency spectrum. 
Finally, in Sec.~\ref{sec:Conclusions}, we conclude. 
We use geometrized units in which $G=c=1$ and the Einstein summation convention
throughout this paper.

\section{WKB Approximation for the Quasinormal-Mode Spectrum of Kerr Black
Holes} 
\label{sec:WKB}
 
In this section, we will derive expressions for the frequencies of quasinormal 
modes of Kerr black holes using the WKB approximation.  
We will need to compute the real and imaginary parts to an accuracy of $O(1)$ 
in terms of $l \gg 1$, which implies that we must calculate $\omega_R$ to 
leading and next-to-leading order and $\omega_{I}$ to leading order.  
Here, we will focus on obtaining an analytic approximation to the frequency 
spectrum, and we will leave the geometrical interpretation of our results until
the next section. 
 
Before specializing our results to the angular and radial Teukolsky equations,
we will review a basic result about the WKB expansion that we will use 
frequently throughout this paper; a more complete discussion of WKB methods 
can be found in \cite{Iyer2}.
Given a wave equation for $\psi(x)$
\begin{equation}
\epsilon^2\frac{d^2 \psi}{d x^2}+U(x) \psi=0 \, ,
\end{equation}
we will expand the solution as $\psi=e^{S_0/\epsilon+S_1+\epsilon S_2+...}$,
where the leading and next-to-leading action variables are given by
\begin{subequations}
\begin{align}
S_0 &= \pm i \int ^x \sqrt{U(x)}dx \, , \\
S_1 &=-\frac{1}{4}\log{U(x)} \, .
\end{align}
\end{subequations}
The formulas above will be the basis for our analysis of the radial and 
angular Teukolsky equations in the next sections.

\subsection{The Teukolsky equations}
\label{sec2b}

Teukolsky showed that scalar, vector, and tensor perturbations of the Kerr 
spacetime all satisfy a single master equation for scalar variables of spin 
weight $s$; moreover, the master equation can be solved by separation of 
variables~\cite{Teukolsky}. 
We will use $u$ to denote our scalar variable, and we will separate this 
scalar wave as 
\begin{equation}
\label{eq6}
u(t,r,\theta,\phi) = e^{-i \omega t }e^{i m \phi}u_r(r)u_{\theta}(\theta) \, .
\end{equation}
Then, at the relevant order in $l \gg 1$, the angular equation for 
$u_\theta(\theta)$ can be written as
\begin{align}
\label{AngTeuk:theta}
\frac{1}{\sin \theta}\frac{d}{d\theta}\left[\sin{\theta}
\frac{d u_{\theta}}{d \theta}\right]+
\left[a^2\omega^2\cos^2{\theta}-\frac{m^2}{\sin^2{\theta}}+A_{lm}\right]
u_{\theta}=0 \,,
\end{align}
where $A_{lm}$ is the angular eigenvalue of this equation. 
The equation obeyed by the radial function $u_r(r)$ is
\begin{subequations}
\begin{equation}
\label{eqr}
\frac{d^2 u_r}{d r^2_*}+\frac{K^2-\Delta\lambda^0_{l m}}{(r^2+a^2)^2} u_r=0 \,,
\quad
\frac{d}{dr_*} \equiv \frac{\Delta}{r^2+a^2}\frac{d}{dr}
\end{equation}
with
\begin{align}
K&=-\omega(r^2+a^2)+am \,, \\
\lambda^0_{l m} &=A_{lm}+a^2\omega^2-2am\omega \,, \\
\Delta &= r^2 - 2 M r +a^2 \,.
\label{eqexplan}
\end{align}
\end{subequations}
Here we have used the facts that $\omega_R \sim O(l)$, $\omega_I \sim O(1)$, 
$m \sim O(l)$ to drop terms that are of higher orders in the expansion than 
those that we treat.  
Note that the spin $s$ of the perturbation no longer enters into these 
equations after neglecting the higher-order terms.

\subsection{The angular eigenvalue problem}

We will first find an expression for $A_{lm}$ in terms of $\omega$, $l$, and
$m$, by analyzing the angular equation in the WKB approximation. 
By defining 
\begin{equation} 
x = \log\left(\tan\frac{\theta}{2}\right)
\end{equation} 
and $dx = d\theta/\sin{\theta}$, we can write the angular equation as 
\begin{subequations}
\begin{equation}
\label{equtheta}
\frac{d^2u_{\theta}}{d x^2}+V^\theta u_{\theta}=0 \,,
\end{equation}
where
\begin{equation}
\label{eq:Vtheta}
V^\theta = a^2\omega^2\cos^2{\theta}\sin^2{\theta}-m^2+A_{lm}\sin^2{\theta} \,.
\end{equation}
\end{subequations}
When written in this form, it is clear that $u_\theta$ must satisfy a boundary
condition that it be 0 as $x \rightarrow \pm \infty$ (which corresponds to 
$\theta \rightarrow 0, \pi$). 
Furthermore, the angular equation is now in a form that is amenable to a WKB 
analysis (which will be the subject of the next part).

First, however, we outline how we will perform the calculation.
Because the frequency $\omega = \omega_R - i \omega_I$ is complex, the angular 
eigenvalue $A_{lm}$, a function of $\omega$, must also be complex.
We will write
\begin{equation}
A_{lm} = A^R_{lm} + i A^I_{lm} \,,
\end{equation}
to indicate the split between real and imaginary parts.   
We will treat a real-valued $\omega =\omega_R$ in the first part of this 
section, and, therefore, a real-valued $A^R_{lm}(\omega_R)$; we shall account 
for $-i \omega_I$ by including it as an additional perturbation in the next 
part of this section.

\subsubsection{Real part of $A_{lm}$ for a real-valued $\omega$}

For $\omega_R \in \mathbb{R}$, we will compute the eigenvalues 
$A_{lm}^R(\omega_R)$, of Eq.~\eqref{equtheta} for standing-wave solutions that 
satisfy physical boundary conditions.  
At the boundary, $\theta = 0,\pi$ (or $x = \mp \infty$) the potential 
satisfies $V^\theta = -m^2$ independent of the value of $A_{lm}^R$; this 
implies that the solutions to Eq.~\eqref{equtheta} behave like decaying 
exponential functions at these points (i.e., the wave does not propagate).
For there to be a region where the solutions oscillate (i.e., where the wave
would propagate), $A_{lm}$ must be sufficiently large to make $V^\theta >0$ in 
some region. 
Depending on the relative amplitudes of $A_{lm}$ and $a^2\omega^2$, $V^\theta$ 
either has one maximum at $\theta=\pi/2$ (when $ A_{lm} \ge a^2\omega^2$), or 
two identical maxima at two locations at symmetrically situated around 
$\theta=\pi/2$ (when $ A_{lm} < a^2\omega^2$).  
It turns out that the region where the maximum of $V^\theta >m^2$ is centered
around $\pi/2$; therefore, all solutions fall into the former category rather 
than the latter. 

The length scale over which the function $u_\theta$ varies is 
$1/\sqrt{V^\theta}$, and the WKB approximation is valid only if the potential
$V^\theta$ does not vary much at this scale.
Therefore, to use the WKB approximation, we require that 
\begin{equation}
\left|\frac{1}{\sqrt{V^\theta}}\frac{dV^\theta}{d\theta}\right| \ll |V^\theta| 
\,.
\end{equation}
This condition applies regardless of the sign of $V^\theta$.  
Empirically, we find this condition to hold for $V^\theta$ in Eq.\
\eqref{equtheta}, except around points at which $V^\theta =0$.  
We will refer to these as {\it turning points}, and they can be found by 
solving for the zeros of the potential,
\begin{equation}
a^2\omega_{R}^2\cos^2{\theta_\pm}\sin^2{\theta_\pm}-m^2+A^R_{lm} \sin^2\theta_\pm=0
\, ,
\end{equation}
which are given by
\begin{equation}
\label{eq:TurningPoints}
\sin^2\theta_\pm =\frac{2 m^2}{A_{lm}+a^2 \omega_{lm}^2
\mp\sqrt{(A_{lm}+a^2 \omega_{lm}^2)^2 +4m^2}} \,.
\end{equation}
Using the leading and next-to-leading WKB approximation, we can write the 
solution to the wave equation in the propagative region, $x_- < x < x_+$, as
\begin{equation}
\label{eq:uthetaWKB}
u_\theta(x)  =
\frac{a_+ e^{i \int^x_0 dx' \sqrt{V^\theta(x')}} + a_- e^{-i \int^x_0 dx' \sqrt{V^\theta(x')}}}{\left[V^\theta(x)\right]^{1/4}} \,,
\end{equation}
where $a_\pm$ are constants that must be fixed by the boundary conditions that
the solution approach zero at $\theta=0,\pi$. 
For $x>x_+$, we find
\begin{subequations}
\begin{equation}
\label{eq:uplusthetaWKB}
u_\theta(x)  =
\frac{c_+ e^{ -\int_{x_+}^{x} dx' \sqrt{-V^\theta(x')}}}
{\left[V^\theta(x)\right]^{1/4}} \, ,
\end{equation}
and $x<x_-$,
\begin{equation}
\label{eq:uminusthetaWKB}
u_\theta(x)  =
\frac{c_- e^{- \int_{x}^{x_-} dx' \sqrt{-V^\theta(x')}}}
{\left[V^\theta(x)\right]^{1/4}} \, ,
\end{equation}
\end{subequations}
with $c_\pm$ also being constants set by the boundary conditions. 
Note that outside of the turning points, we have only allowed the solution that
decays towards $x \rightarrow \pm \infty$. 

Around the turning points $x_\pm$, the WKB approximation breaks down, but 
$u_\theta$ can be solved separately by using the fact that 
$V_\theta (x\sim x_\pm) \propto x - x_\pm$. 
Solutions obtained in these regions can be matched to Eqs.\
\eqref{eq:uthetaWKB}--\eqref{eq:uminusthetaWKB}; the matching condition leads 
to the Bohr-Sommerfeld quantization condition 
\begin{equation}
\label{Aeq}
\int_{\theta_-}^{\theta_+}
d\theta \sqrt{a^2\omega_{R}^2 \cos^2\theta -\frac{m^2}{\sin^2\theta}+A_{lm}^R} 
=\left(L-|m|\right)\pi\,.
\end{equation}
Here we have defined
\begin{equation}
L \equiv l+\frac{1}{2} \, ,
\end{equation}
which will be used frequently throughout this paper.  
The limits of the integration are the values of $\theta$ where the integrand 
vanishes [the turning points of Eq.\ (\ref{eq:TurningPoints})].

If we define 
\begin{equation}
\mu \equiv \frac{m}{L}\,,\;\;
\alpha_R(a,\mu) \equiv \frac{A^R_{lm}}{L^2}\,, \; \;
 \Omega_R(a,\mu)\equiv \frac{\omega_R}{ L}\,,
\end{equation}
then all three of these quantities are $O(1)$ in our expansion in $L$. 
From these definitions, we can re-express the limits of integration as
\begin{equation}
\sin^2\theta_\pm =\frac{2\mu^2}{\alpha+a^2 \Omega^2  
\mp \sqrt{(\alpha+a^2 \Omega_R^2 )^2+4\mu^2}} \,,
\end{equation} 
and the integral as 
\begin{equation}
\int_{\theta_-}^{\theta_+}  d\theta \sqrt{\alpha_R -\frac{\mu^2}{\sin^2\theta}+ a^2\Omega^2 \cos^2\theta} =(1-|\mu|)\pi \,.
\end{equation}

For each set of quantities $(\alpha_R,\mu,\Omega_R)$, we can express $\alpha_R$
as an implicit function involving elliptic integrals; however, if we treat 
$a \Omega_R$ as a small parameter, then the first two terms in the expansion
are
\begin{equation}
\label{AlmRapp}
\alpha_R \approx 1 -\frac{a^2 \Omega_R^2}{2} \left (1-|\mu|^2\right ) \,.
\end{equation}
We derive and discuss this approximation in greater detail in Appendix
\ref{sec:BohrSommerfeldAp}.
For $a=0$, we note that this is accurate with a relative error of $O(1/L^2)$, 
because for a Schwarzschild black hole 
\begin{equation}
A_{lm}^{\rm Schw} =l(l+1)-s(s+1)\,.
\end{equation}
As we will confirm later in Figs.\ \ref{fig:error} and \ref{fig:error2}, 
Eq.~\eqref{AlmRapp} is an excellent approximation even for highly spinning 
black holes.  

To understand intuitively why the approximation works so well, we will focus on
corotating modes (i.e., those with positive and large $m$, or $\mu$ near 
unity), which have the highest frequencies and, therefore, the largest possible 
values for $a\Omega_R$.  
For a fixed value of $(l,m)$, $\omega_R$ is a monotonically increasing function
of $a$, and 
\begin{equation}
\omega_R^{lm} (a) \le \omega_R^{lm} (a=M) = m\Omega_H^{a=1}=\frac{m}{2M}\,.
\label{eq:FreqBound}
\end{equation}
In setting this upper bound, we have used the result that the low-overtone QNM 
frequencies approach $m \Omega_H$ for $m>0$ and for extremal black holes (first
discussed by Detweiler \cite{Detweiler}); we have also used $\Omega_H$ to 
denote the horizon frequency of the Kerr black hole,
\begin{equation}
\Omega_H = \frac{a}{2 M r_+}\,,
\end{equation}
and $r_+$ to indicate the position of the horizon [note that $r_+(a=M) = M$].
Normalizing Eq.\ \ref{eq:FreqBound} by $L$, we find
\begin{equation}
a\Omega_R \leq (\mu/2) (a/M) \leq 1/2\,.
\end{equation}
Even for the upper bound $a\Omega_R=1/2$, the relative accuracy of 
Eq.~\eqref{AlmRapp} is still better than $0.2\%$.

\subsubsection{Complex $A_{lm}$ for a complex $\omega$} 
\label{sec:ComplexAlm}

To solve for the next-to-leading-order corrections to $A_{lm}$, we must compute
the imaginary part $A^I_{lm}$. 
Because $\omega_I \ll \omega_R$, when we allow $\omega =\omega_R - i\omega_I$ 
to be a complex number in the angular eigenvalue problem \eqref{AngTeuk:theta},
we can treat the term linear in $\omega_I$ as a perturbation to the angular 
equation. 
Using the perturbation theory of eigenvalue equations, we find that
\begin{equation}
A_{lm}^I = -2a^2\omega_R\omega_I \langle\cos^2\theta\rangle \, ,
\label{eq:ImagAlm}
\end{equation} 
where 
\begin{align}
\langle\cos^2\theta\rangle &= \frac{\displaystyle\int \cos^2\theta |u_\theta|^2\sin\theta d\theta}{\displaystyle\int |u_\theta|^2\sin\theta d\theta}  \nonumber\\
&=\frac{\displaystyle \int_{\theta_-}^{\theta_+} \frac{\cos^2\theta}{\sqrt{a^2\omega_{R}^2 \cos^2\theta -\frac{m^2}{\sin^2\theta}+A_{lm}^R}}  d\theta}{\displaystyle \int_{\theta_-}^{\theta_+} \frac{1}{\sqrt{a^2\omega_{R}^2 \cos^2\theta -\frac{m^2}{\sin^2\theta}+A_{lm}^R}} d\theta} \,.
\end{align}
By taking the derivative of both sides of the Bohr-Sommerfeld condition  
\eqref{Aeq} with respect to the variable $z = a \omega_R$ and by treating 
$A_{lm}$ as a function of $z$, we can rewrite the above expression as
\begin{align}
\langle\cos^2\theta\rangle &=\left.
- \frac{1}{2z}\frac{\partial A_{lm}^R(z)}{\partial z}\right|_{z=a \omega_R }\,.
\end{align}
Substituting this expectation value into Eq.\ (\ref{eq:ImagAlm}), we find
\begin{equation}
\label{almI}
A_{lm}^I = a \omega_I 
\left[\frac{\partial A_{lm}^R(z)}{\partial z}\right]_{z=a \omega_R }\,.
\end{equation}
Equation (\ref{almI}) defines a numerical prescription for computing 
$A_{lm} =A_{lm}^R+iA_{lm}^I$. 
This approach is quite natural: as $\omega$ becomes complex, $A_{lm}$ is the 
analytic function whose value on the real axis is given by $A_{lm}^R$. 
The approximate formula \eqref{AlmRapp}, therefore, becomes
\begin{subequations}
\begin{equation}
\label{Almfullapp}
A_{lm}\approx L^2-\frac{a^2\omega^2}{2}
\left [1-\frac{m^2}{L^2}\right ] \,,
\end{equation}
or
\begin{equation}
\alpha \approx 1 -\frac{a^2 \Omega^2}{2}
\left (1-|\mu|^2\right ) \,,
\end{equation}
\end{subequations}
for a complex frequency $\omega$.

\subsection{The radial eigenvalue problem}

Now that we have solved for the angular eigenvalues $A_{lm}$ in terms of 
$\omega$, we turn to the radial Teukolsky equation. 
From Eq.~\eqref{eqr}, we see that the radial equation is already in the form
\begin{subequations}
\begin{equation}
\frac{d^2 u_r}{dr_*^2} +V^r u_r=0 \, ,
\label{eq:radialTeuk}
\end{equation}
if we define
\begin{equation}
V^r (r,\omega)=\frac{[\omega(r^2+a^2)- ma ]^2 -\Delta  \left[A_{lm}(a \omega) +a^2\omega^2 -2 m a \omega\right] }{(r^2+a^2)^2} \,.
\end{equation} 
\end{subequations}
Note here that $V^r$ is an analytic function of $\omega$, and that it is 
real-valued when $\omega$ is real. 

In general, the WKB approximant for $u_r$ is given at leading order by
\begin{equation}
\label{WKBradial}
u_r = b_+ e^{i\int^{r_*} \sqrt{V^r(r_*')} dr_*'}+b_- e^{-i\int^{r_*} \sqrt{V^r(r_*')} dr_*'} \,,
\end{equation} 
although in order to obtain a mode which is outgoing at 
$r_*\rightarrow +\infty$ (the same as $r\rightarrow \infty$) and ingoing at 
$r_*\rightarrow -\infty$ ($r\rightarrow r_+$), we must have
\begin{subequations}
\begin{equation}
u_r = b_+ e^{i\int^{r_*} \sqrt{V^r(r_*')}} dr_*' \, ,
\end{equation}
for the region containing $r \rightarrow +\infty$, and 
\begin{equation}
u_r = b_- e^{-i\int^{r_*} \sqrt{V^r(r_*')} }dr_*' \, ,
\end{equation}
\end{subequations}
for the region containing $r_* \rightarrow -\infty$. 
Intuitively speaking, a solution to Eq.\ (\ref{eq:radialTeuk}) will satisfy 
the asymptotic behavior above if $V^r \approx 0$ around a point $r=r_0$, and 
$V_r >0$ on both sides.  
Then, the WKB expansion \eqref{WKBradial} is valid in the two regions on both  
sides of $r=r_0$, and the solution in the vicinity of $r_0$ must be obtained
separately by matching to the WKB approximation.
The matching will constrain the frequency, thereby giving a method to determine
$\omega$.  
A detailed calculation of this procedure has been carried out by Iyer and Will
\cite{Iyer2} to high orders in the WKB approximation; the only difference 
between our calculation and their result at lower orders comes from the more 
complex dependence of $V^r$ on $\omega$ in our case (particularly because 
$A_{lm}$ depends on $\omega$ in a more involved way).

\subsubsection{Computing $\omega_R$}

From Iyer and Will~\cite{Iyer2}, the conditions at the leading and 
next-to-leading order that must be solved to find $\omega_R$ are
\begin{equation}
\label{Vreq}
V^r(r_0,\omega_R)=\left.\frac{\partial V^r}{\partial r}\right|_{(r_0,\omega_R)}
=0 \, .
\end{equation}
After a short calculation, these conditions can be expressed as 
\begin{subequations}
\begin{align}\label{v}
\Omega_R &= \frac{ \mu a}{r_0^2+a^2} \pm \frac{\sqrt{\Delta(r_0)}}{r_0^2+a^2}\beta(a \Omega_R) \,, \\
\label{dvdr}
0 &=\frac{\partial}{\partial r}\left[\frac{\Omega_R (r^2+a^2)- \mu a}{\sqrt{\Delta(r)}} \right]_{r=r_0} \,,
\end{align}
\end{subequations}
where we have defined
\begin{subequations}
\begin{align}
\label{blmz}
\beta(z) &=\sqrt{\alpha (z)+z^2 -2 \mu z}   \\
&\approx 
\sqrt{1+\frac{z^2}{2}-2 \mu z+\frac{\mu^2 z^2}{2 }} \,.
\label{blmz2}
\end{align}
\end{subequations}
In deriving Eq.\ (\ref{dvdr}), we have used the fact that at $r>r_+$, 
$(r^2+a^2)^2/\Delta$ is a monotonically increasing function, and, therefore
the extrema of $V^r$ are the same as those of $V^r(r^2+a^2)^2/\Delta$; 
we then also used the fact that the quantity within the square brackets in 
Eq.\ (\ref{dvdr}) is always nonzero at points at which $V^r=0$.

One method of jointly solving Eqs.\ (\ref{v}) and (\ref{dvdr}) would be to 
use Eq.~\eqref{dvdr} to express $\Omega_R$ in terms of $r_0$
\begin{equation}
\label{omegarr0}
\Omega_R =\frac{(M -r_0) \mu a}{(r_0-3M)r_0^2 +(r_0+M)a^2}\,,
\end{equation}
and then insert this into Eq.~\eqref{v} to obtain $r_0$; finally $\Omega_R$ can
be obtained by substituting this $r_0$ back into Eq.~\eqref{omegarr0}.
If we use the approximate formula \eqref{blmz2} in this process, the equation 
for $r_0$ becomes a sixth-order polynomial in $x = r_0/M$, the roots of which 
can be found efficiently. 
For conveninece, we write this polynomial here
\begin{align}
\label{poly}
2x^4(x-3)^2 +4x^2[(1-\mu^2)x^2-2x-3(1-\mu^2)](a/M)^2 \nonumber\\
+(1-\mu^2)[(2-\mu^2)x^2 + 2(2+\mu^2)x + (2-\mu^2)](a/M)^4\,.
\end{align}
For each pair $(\mu,a/M)$, there are in general two real roots for $x$, which 
correspond to the two possible values of $r_0/M$ (and the two real frequencies 
with opposite signs). 

Note that the procedure above will not work when $m=0$ (when both the numerator
and denominator of Eq.~\eqref{omegarr0} vanish).  
In this case, we can directly require that 
\begin{equation}
\label{r0}
(r_p-3M)r_p^2 +(r_p+M)a^2 =0 \,.
\end{equation}
The solution, $r_p$ can be inserted into Eq.\ (\ref{v}) and the result can be
expressed in terms of elliptic integrals
\begin{equation}
\Omega_R(a, \mu=0) = \pm\frac{\pi \sqrt{\Delta(r_p)}} 
{(r_p^2+a^2)\mathrm{EllipE}
\left[\displaystyle\frac{a^2\Delta(r_p)}{(r_p^2+a^2)^2}\right]} \,,
\end{equation}
where $\mathrm{EllipE}$ denotes an elliptic integral of the second kind.
Here we have used the subscript $p$ for this special case, because this mode 
will turn out to correspond to polar orbits. 

\begin{figure}[t!]
\includegraphics[width=0.95\columnwidth]{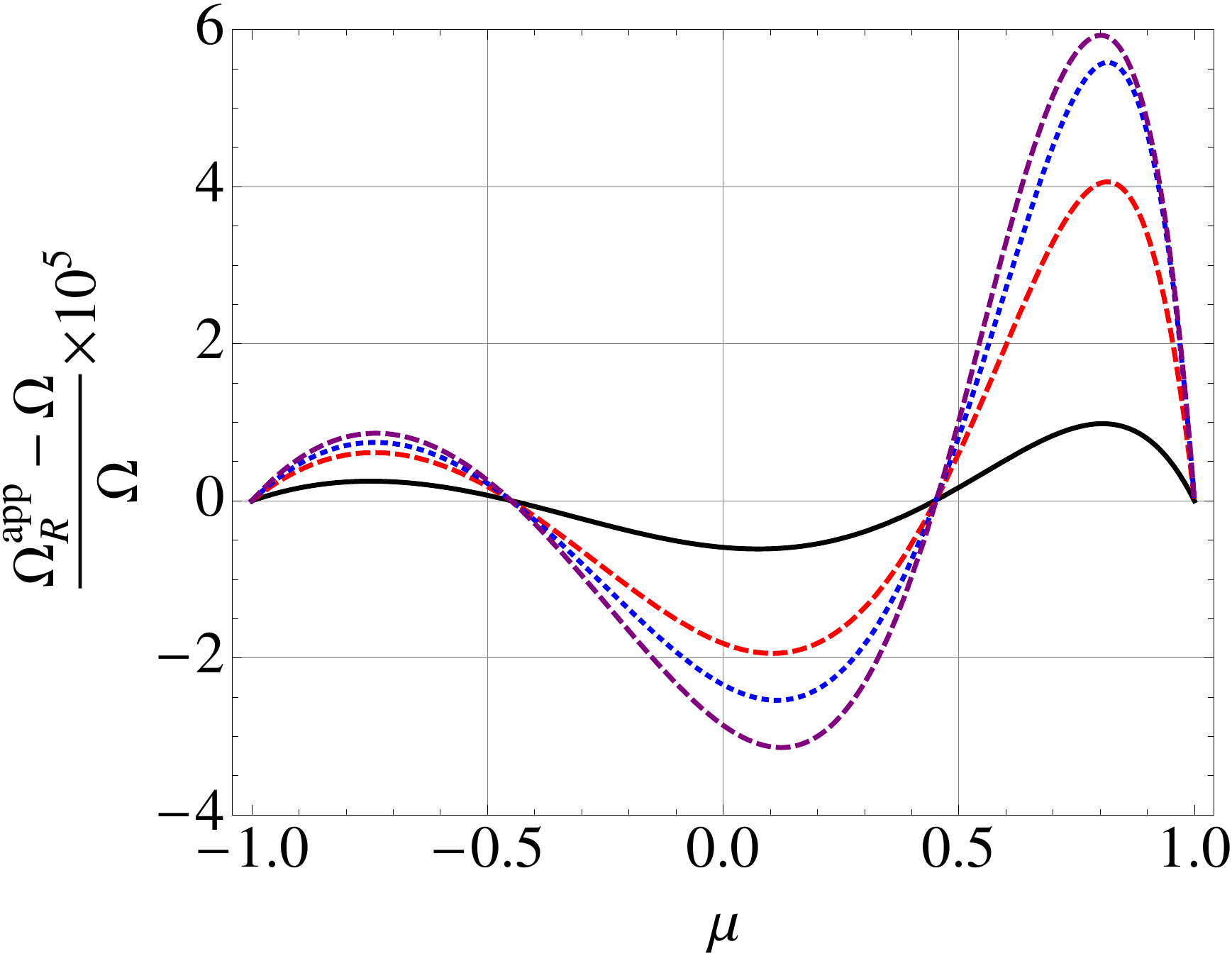}
\caption{Difference in $\Omega_R(a,\mu)$ [Eq.~\eqref{omegarr0}] that arises 
from using the approximate formula for $A_{lm}$ [Eq.~\eqref{Almfullapp}] as 
opposed to the exact formula.
Here $a/M=0.7$, $0.9$,$ 0.95$, and $0.99$ correspond to black solid, red 
dashed, blue dotted, and purple long-dashed curves, respectively.
The quantity plotted on the vertical axis has been scaled by $10^5$.}
\label{fig:error}
\end{figure}

\begin{figure}[t!]
\includegraphics[width=0.95\columnwidth]{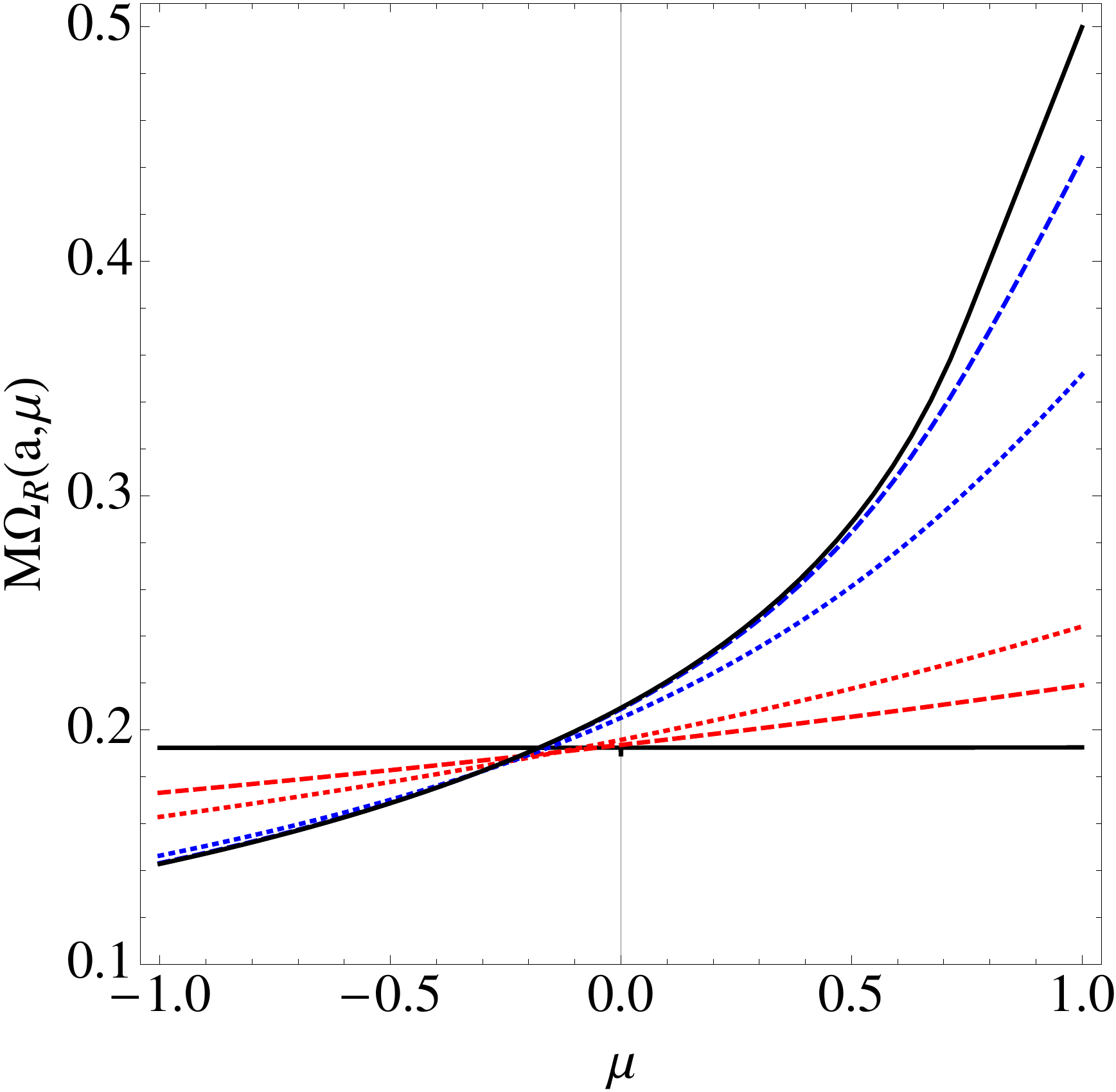}
\caption{Real part of the QNM spectra from the WKB approximation.  Black solid 
curves show $\Omega$ for $a/M=0$ (the flat curve) and $a/M=1$ (the curve that 
increases towards 0.5); red (light gray) dashed and dotted curves show 
$a/M=0.3$ and $0.5$, while blue (dark gray) dotted and dashed curves show 
$a/M=0.9$ and $0.99$.}
\label{fig:Omega}
\end{figure}

We plot in Fig.~\ref{fig:error} the relative error in $\Omega_R$ that comes
from using the approximate expression for $A_{lm}$ [Eq.~\eqref{Almfullapp}] 
rather than the exact Bohr-Sommerfeld condition. 
The error is always less than $\sim  10^{-4}$ (we scale the quantity plotted on
the vertical axis by $10^5$), and therefore, we will use the approximate 
expression for $A_{lm}$ for the remaining calculations involving $\Omega_R$ 
throughout this paper.
In Fig.~\ref{fig:Omega}, we plot $\Omega_R$ for $a/M=0$, $0.3$, $0.5$, $0.9$,
$0.99$, and $1$ (the flat curve corresponds to $a/M=0$, and those with 
increasing slopes are the increasing values of $a/M$).  
While for low values of $a/M$ below $\sim 0.5$, $\Omega_R$ depends roughly
linearly upon $\mu$, for higher values of $a/M$ (and for $\mu>0$), $\Omega_R$ 
grows more rapidly than linearly.  
For $a/M=1$, $\Omega_R\rightarrow 1/2$ when $\mu\rightarrow 1$, as anticipated.

\subsubsection{Computing $\omega_I$}

At leading order, the imaginary part $\omega_I$ can be calculated using the 
same procedure set forth by Iyer and Will~\cite{Iyer2}.
They find that
\begin{align}
\label{omegai}
\omega_I & =-(n+1/2) \frac{\sqrt{2\left(\frac{d^2 V^r}{dr_*^2}\right)_{r_0,\omega_R}}}{\left(\frac{\partial V^r}{\partial \omega}\right)_{r_0,\omega_R}} \, ,
\nonumber\\
& =-(n+1/2)\Omega_I(a,\mu) \, .
\end{align}
In our calculation, we must also take into account that $V^r$ also depends on 
$\omega$ through the angular eigenvalue's dependence on $\omega$. 
If we use the approximate formula for $\alpha$, we obtain a reasonably compact
expression for $\Omega_I$:
\begin{widetext}
\begin{align}
\Omega_I &=\Delta(r_0)\frac{\sqrt{4(6r_0^2 \Omega_R^2 - 1) + 2 a^2 \Omega^2_R(3 - \mu^2)}}{2 r_0^4 \Omega_R  - 4 a M r_0 \mu + a^2 r_0 \Omega_R [r_0(3 - \mu^2) + 2 M (1 + \mu^2)] + a^4 \Omega_R(1 - \mu^2)} \,.
\end{align}
\end{widetext} 

\begin{figure}[t!]
\includegraphics[width=0.95\columnwidth]{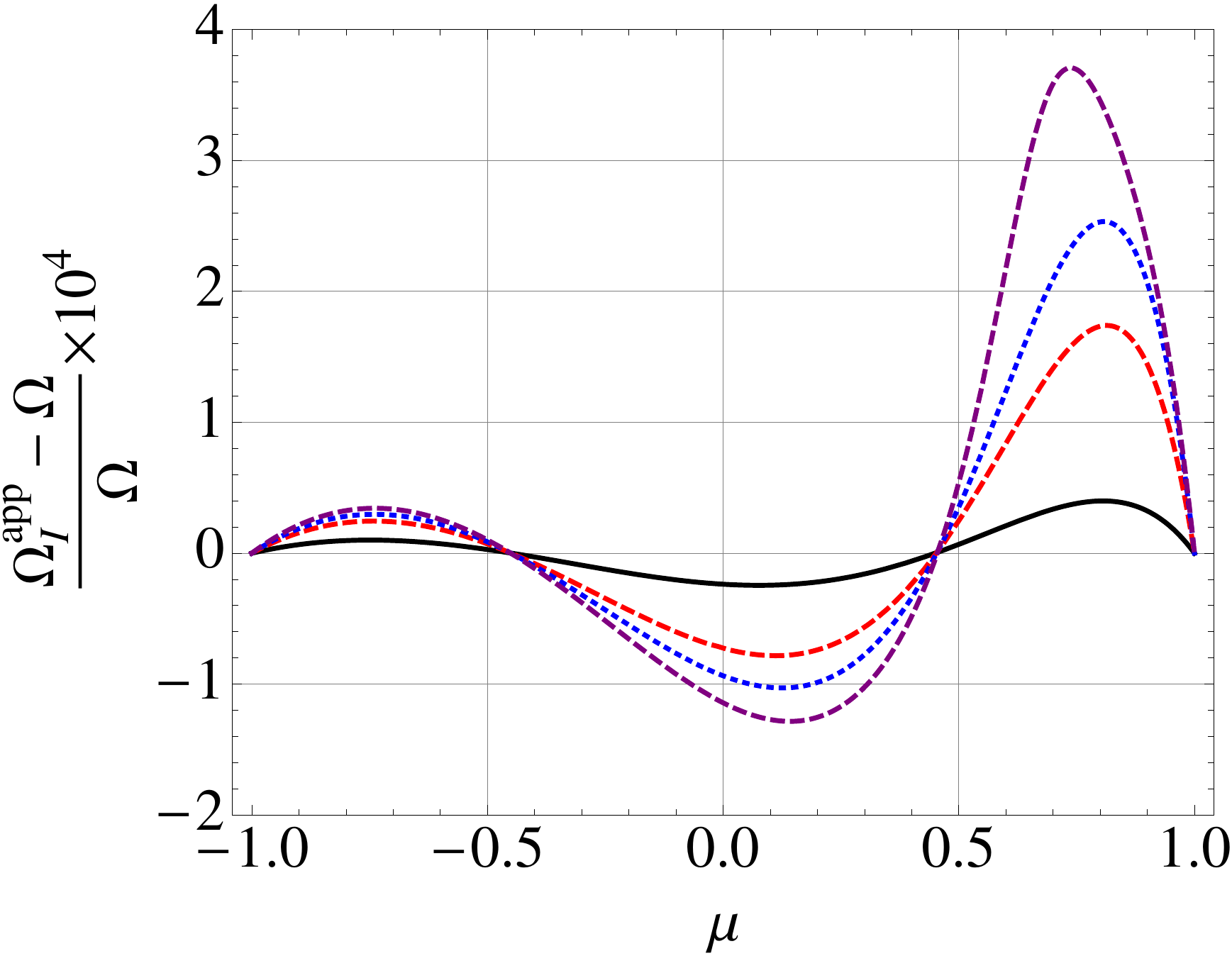}
\caption{Difference in $\Omega_I(a,\mu)$ [Eq.~\eqref{omegai}] from using the 
approximate formula for $A_{lm}$ [Eq.~\eqref{Almfullapp}] rather than the exact
formula.  
Here $a/M=0.7$, $0.9$, $0.95$, and $0.99$ correspond to black solid, red 
dashed, blue dotted, and purple long-dashed curves, respectively.
We scale the quantity plotted along the vertical axis by $10^4$ in this 
figure.}
\label{fig:error2}
\end{figure}

\begin{figure}[t!]
\includegraphics[width=0.95\columnwidth]{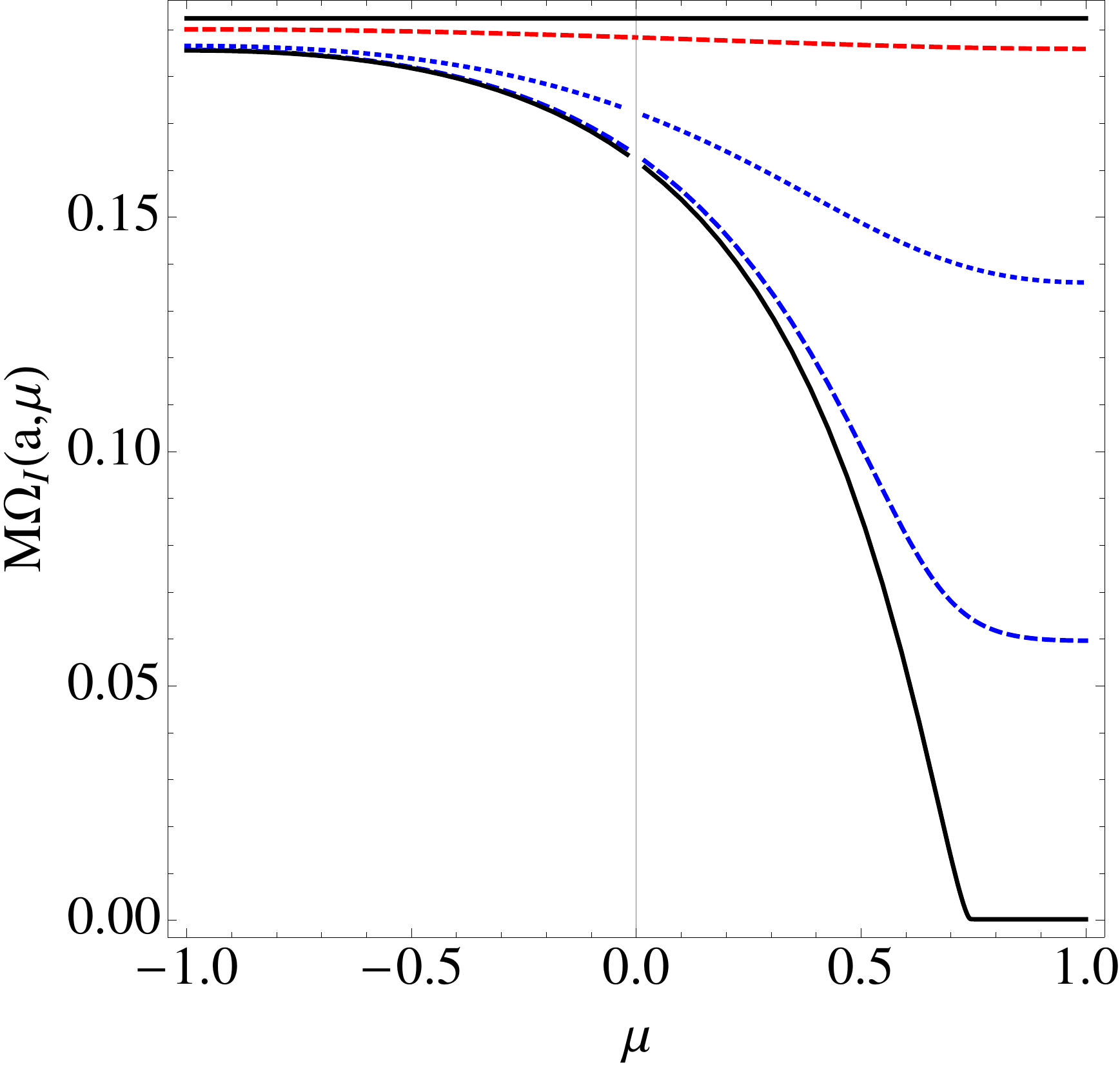}
\caption{Imaginary part of the QNM spectrum computed in the WKB approximation.
Black solid curves show $\Omega_I$ for $a/M=0$ (again the flat curve) and 
$a/M=1$ the curve that decreases and heads to zero.
The red dashed curve shows $a/M=0.5$, while blue dotted and dashed curves show 
$a/M=0.9$ and $0.99$, respectively.  
For $a/M=1$, modes with $\mu \gtrsim 0.74$ approach zero (modes that do no 
decay), while others still decay.}
\label{fig:OmegaI}
\end{figure}

In Fig.~\ref{fig:error2}, we plot the relative error in $\Omega_I$ from using 
the approximate formula for $A_{lm}$ identically to that in Fig.\ 
\ref{fig:error} (although here we scale the quantity plotted on the vertical 
axis by $10^4$).
Because the error is always less than $\sim 10^{-3}$, we will use the 
approximate expression for $A_{lm}$ for computing $\Omega_I$ in the remainder
of this paper.
In Fig.~\ref{fig:OmegaI}, we plot $\Omega_I(a,\mu)$ for several values of 
$a/M$ (the same as those in Fig.\ \ref{fig:Omega}, though not $a/M=0.3$).  
The curve for $a/M=0$ is flat, and those with larger spins have more rapidly
decreasing slopes for increasing values of $a/M$.
It is interesting to note that in the limit $a\rightarrow 1$, $\Omega_I$ 
becomes zero for values of $\mu$ in a finite interval $0.74\lesssim\mu\leq 1$ 
(not only for $\mu = 1$ does $\Omega_I$ vanish).  
We will put forward an explanation for this phenomenon in Sec.\ 
\ref{sec:Spectra}, after we make connections between QNMs and wave propagation 
in the Kerr spacetime. 

\subsection{Accuracy of the WKB approximation}
\label{sec:WKBaccuracy}

\begin{figure*}
\includegraphics[width=0.9\textwidth]{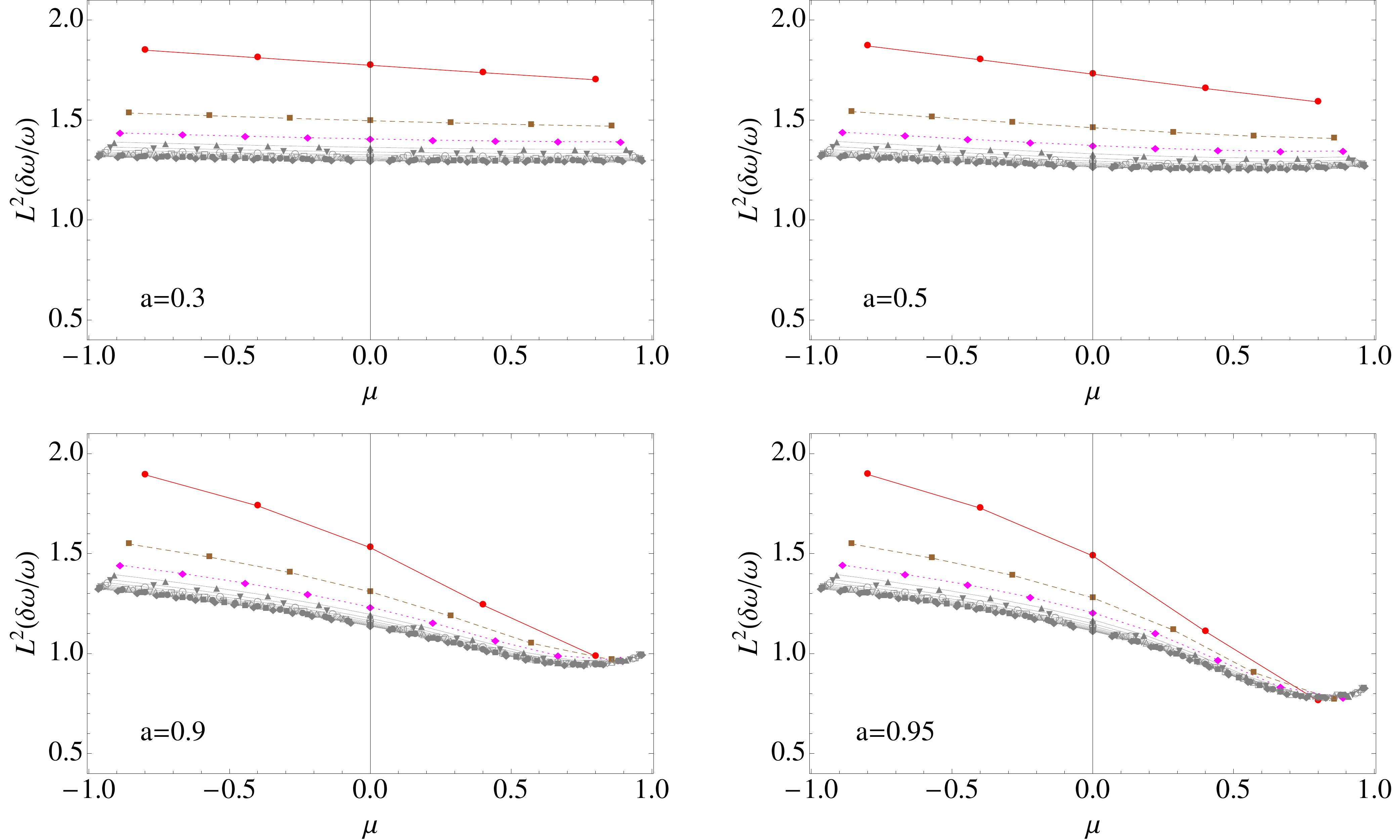}
\caption{Fractional error, $\delta\omega_R/\omega_R$, of the WKB approximation 
to the $s=2$, gravitational-wave, quasinormal-mode spectrum, multiplied by 
$L^2$.
The four panels correspond to four different spins which (going clockwise from
the top left) are $a/M=0.3$, $0.5$, $0.95$, and $0.9$.
Errors for $l=2,3,4$ are highlighted as red solid, brown dashed, and pink 
dotted lines, while the rest ($l=5,\ldots,14$) are shown in gray.
This shows that the relative error approaches the $O(1/L^2)$ scaling quite
quickly.}
\label{fig:compare:s2}
\end{figure*}

\begin{figure*}
\includegraphics[width=0.9\textwidth]{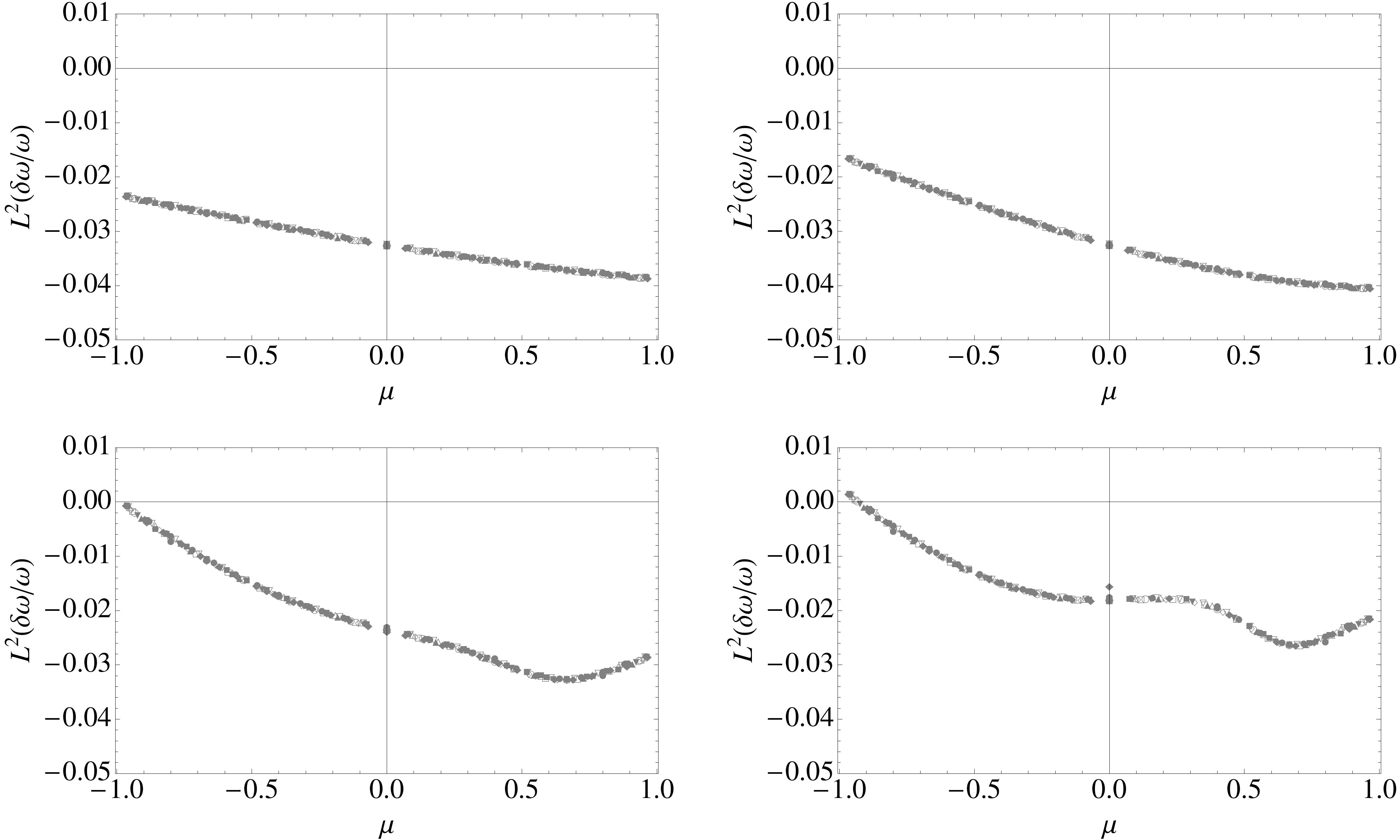}
\caption{Fractional error, $\delta\omega_R/\omega_R$, of the WKB approximation 
to the $s=0$, scalar-wave, quasinormal-mode spectrum, again scaled by $L^2$.
The four panels correspond to the same four spins in Fig.\ 
\ref{fig:compare:s2}.
The points shown in the four panels are for values of $l$ in the range 
$l=2,3,\ldots,14$.
Because all values of $l$ nearly lie on the same curve, the relative error has 
converged at an order $O(1/L^2)$ even for very low $l$.
The overall error is also significantly lower than that for the $s=2$ modes.}
\label{fig:compare:s0}
\end{figure*}

\begin{figure*}
\includegraphics[width=0.9\textwidth]{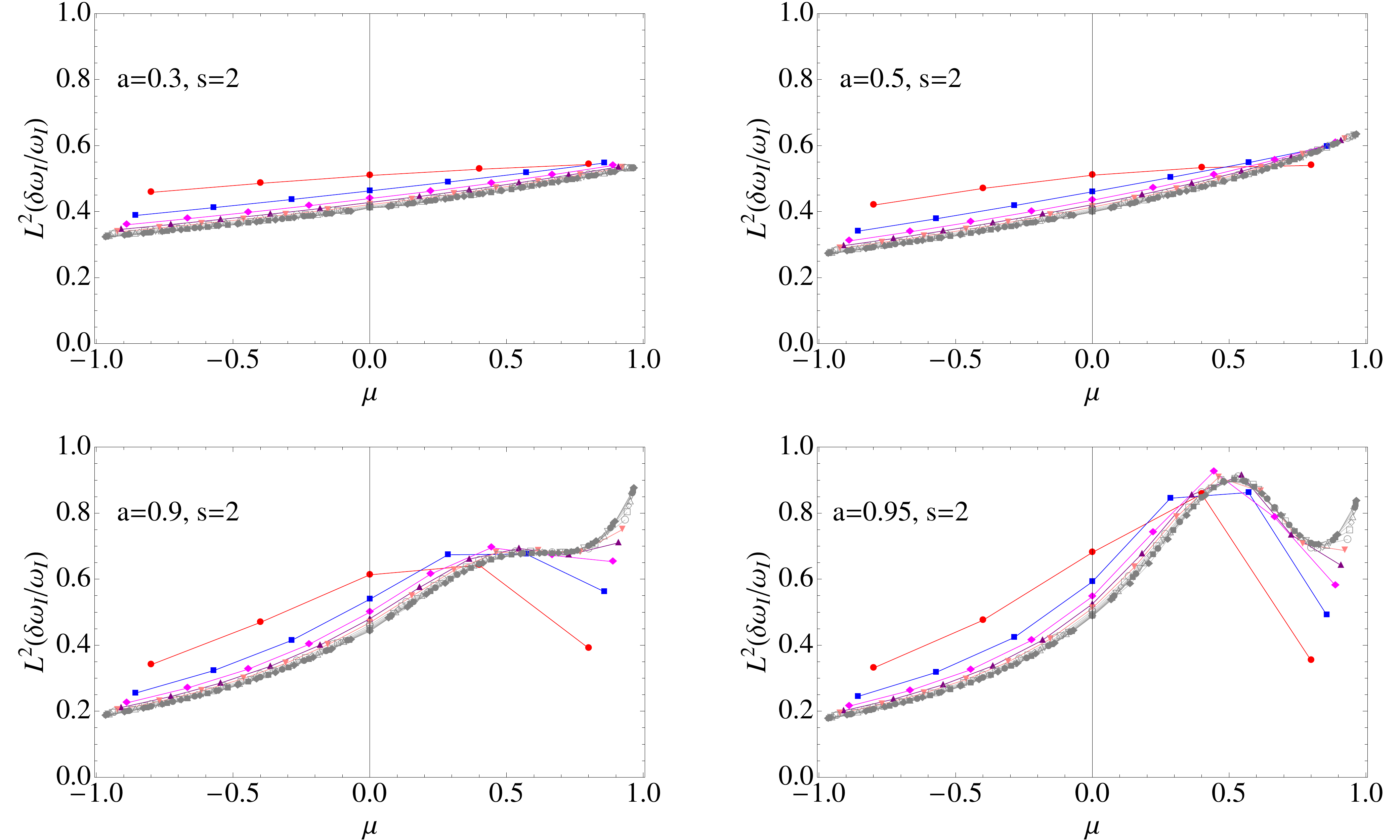}
\caption{Fractional error, $\delta\omega_I/\omega_I$, of the WKB approximation 
to the $s=2$, gravitational-wave, quasinormal-mode spectrum, also scaled by 
$L^2$.  
The panels and the curves are plotted in the same way as in Fig.\ 
\ref{fig:compare:s2}, and the error scales similarly.} 
\label{fig:compare:imags2}
\end{figure*}

\begin{figure*}
\includegraphics[width=0.9\textwidth]{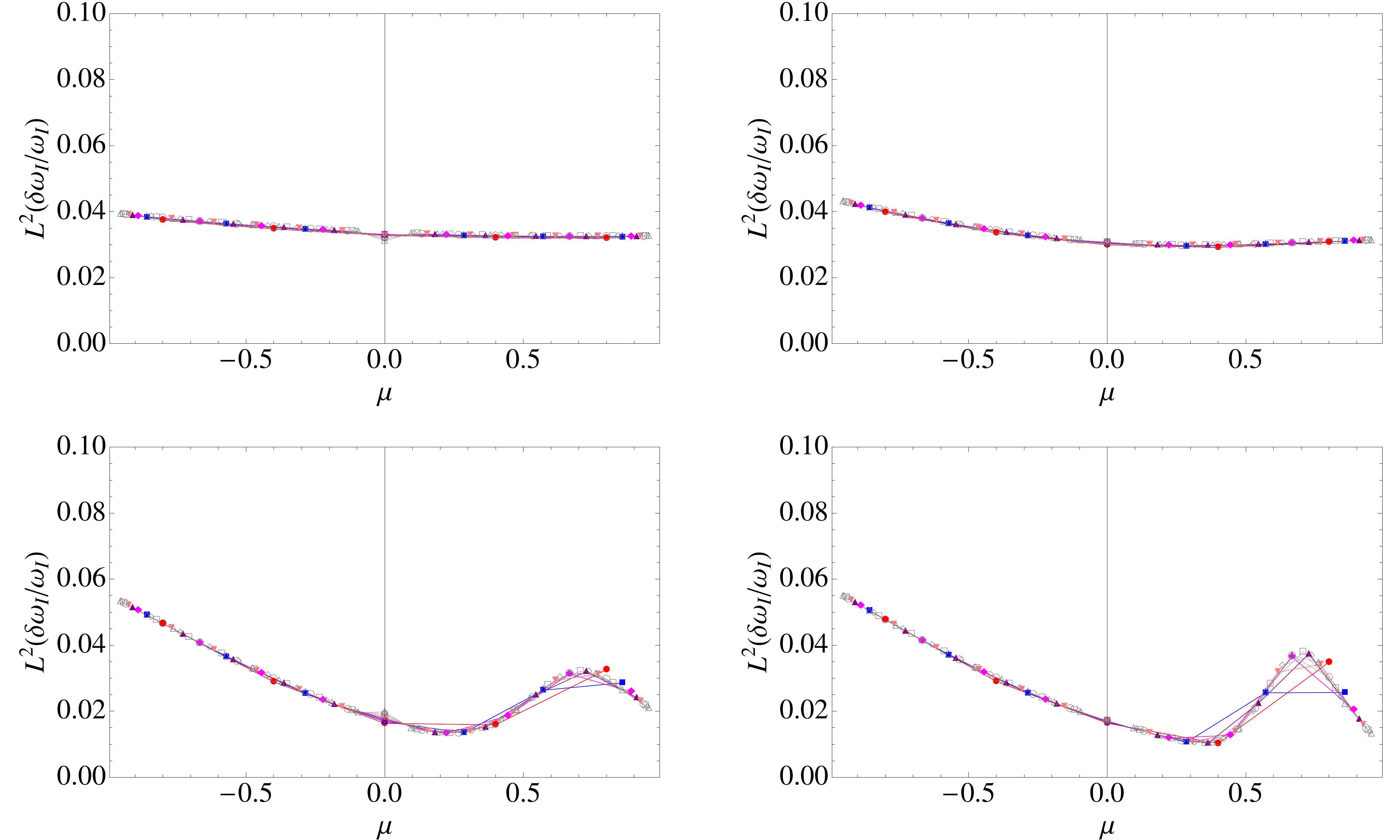}
\caption{Fractional error, $\delta\omega_I/\omega_I$, of the WKB approximation 
to the $s=0$, scalar-wave, quasinormal-mode spectrum, again multiplied by 
$L^2$.
The four panels and the points are shown in the same way as in Fig.\ 
\ref{fig:compare:s0}, and there is a similar rapid convergence of the error.}
\label{fig:compare:imags0}
\end{figure*}

Because we calculated the leading and next-to-leading orders in the WKB 
approximation to $\omega_R$, we expect that the relative error for increasing
$L$ should scale as $O(1/L^2)$.
For the imaginary part, we computed only the leading-order expression, and we 
would expect that the relative error might scale as $O(1/L)$.  
In addition, because at this order of approximation, we do not account for the 
spin of the wave, we anticipate that the error for the gravitational modes may
be larger than those for scalar modes.
In Figs.\ \ref{fig:compare:s2}--\ref{fig:compare:imags0}, we confirm most of
these expectations, but we find the somewhat unexpected result that the 
relative error for the imaginary part also scales as $O(1/L^2)$.

In Fig.~\ref{fig:compare:s2}, we compare the WKB approximation to $\omega_R$ 
with numerical computations of the $s=2$, gravitational-wave, quasinormal-mode 
spectra; specifically, we plot the fractional error against $\mu = m/L$, for 
$l=2,3,\ldots, 14$, and for black holes of spins $a/M=0.3$, $0.5$, $0.9$, and 
$0.95$.  
The relative error clearly converges to $O(L^2)$.
Even for $l=2$, the relative error tends to be $\lesssim 30\%$, and at 
$l \ge 3$ the relative error stays below $\sim 1.5 L^{-2}$ (this means error is
$\lesssim 10\%$ for $l=3$ and higher).

In Fig.~\ref{fig:compare:s0}, we compare the WKB spectra with $s=0$ scalar 
quasinormal-mode spectra, for the same values of $l$ and the same black-hole
spins.
We find a much better agreement.  
For all $l \ge 2$ modes, the relative error stays below 
$4\times 10^{-2} L^{-2}$.  
This suggests that coupling between the spin of the wave (i.e., its tensor 
polarization) and the background curvature of the Kerr black hole is the main 
source of error in our WKB approximation. 

In Figs.~\ref{fig:compare:imags2} and \ref{fig:compare:imags0}, we perform the 
same comparisons as in Figs.\ \ref{fig:compare:s2} and \ref{fig:compare:s0} for
the imaginary part of frequency. 
Surprisingly, we find that for both $s=0$ and $2$, the relative error in 
$\omega_I$ is $O(L^{-2})$.  
For $s=0$, the relative error is $\lesssim 6\times 10^{-2} L^{-2}$, while for 
$s=2$, the error is $\lesssim L^{-2}$. 

With this comparison, we conclude our direct calculation of the QNM spectrum 
by WKB techniques. 
We will discuss additional features of the QNM spectrum in Sec.\
\ref{sec:Spectra}, but before doing so, we will develop a geometric 
interpretation of our WKB results. 
Doing so will help us to develop more intuition about our WKB expressions.

\section{Geometric Optics in the Kerr Spacetime}
\label{sec:GeomOptics}

In this section, we first briefly review the formalism of geometric optics, 
which describes the propagation of waves with reduced wavelengths $\lambdabar$ 
that are much shorter than the spacetime radius of curvature, $R$, and the size
of the phase front, $\mathcal L$. 
In the geometric-optics approximation, the phase of the waves remains constant 
along null geodesics (rays), while the amplitude can be expressed in terms of 
the expansion and contraction of the cross-sectional area of bundles of null 
rays.  
We will then specialize the geometric-optics formalism to the Kerr spacetime, 
and we will write down the most general form of propagating waves in the
geometric-optics approximation.
Using the Hamilton-Jacobi method, we see that the waves' motion can be related
to the null geodesics in the spacetime.
By applying boundary conditions to the approximate wave, we obtain expressions
for the quasinormal-mode waveforms and their corresponding complex frequency 
spectra and angular separation constants, in the eikonal limit.

\subsection{Geometric optics: general theory}

Here we briefly review the geometric-optics approximation to scalar-wave 
propagation (see, e.g., Section 22.5 of Ref.~\cite{MTW} for details).
A massless scalar wave $u$ propagating in curved spacetime satisfies the 
wave equation
\begin{equation}
\label{eq:wave}
g^{\mu\nu}\nabla_{\mu} \nabla_{\nu} u=0\,.
\end{equation}
If we write
\begin{equation}
\label{eik:u}
u = A e^{i\Phi}\,,
\end{equation}
then at leading order in $\lambdabar/\mathcal{L}$, we have
\begin{equation}
\label{eik:Phi}
g^{\mu\nu} k_\mu k_\nu=0\,,\quad k_\mu \equiv \partial_\mu \Phi\,,
\end{equation}
while at next-to-leading order, 
\begin{equation}
\label{eik:A}
2k^\mu \partial_\mu \log A + \nabla_\mu k^\mu=0\,.
\end{equation}
Note that Eq.~\eqref{eik:Phi} also implies that $k^\mu$ is geodesic,
\begin{equation}
\label{eik:k}
k^\mu \nabla_\mu k_\nu = k^\mu \nabla_\mu \nabla_\nu \Phi =
k^\mu  \nabla_\nu  \nabla_\mu \Phi = k^\mu  \nabla_\nu  k_\mu =0\,.
\end{equation}

Equations~\eqref{eik:u}--\eqref{eik:k} encode information about the transport 
of the amplitude $A$ and phase $\Phi$ along a null geodesic (or a {\it ray}).
The phase should be kept constant, because Eq.~\eqref{eik:Phi} states
\begin{equation}
\label{trans:Phi}
k^\mu \partial_\mu\Phi=0 \,,
\end{equation}
while the amplitude is transported along the ray in a manner that depends upon
the propagation of neighboring rays.  
Because the 2D area, $\mathcal A$, of a small bundle of null rays around the 
central ray satisfies the equation 
\begin{equation}
\nabla_\mu k^\mu = k^\mu \partial_\mu \log\mathcal{A} \,,
\end{equation}
it is possible to show from Eq.\ (\ref{eik:A}) that
\begin{equation}
\label{trans:A}
k^\mu\partial_\mu \left( \mathcal{A}^{1/2} A\right)=0\,,
\end{equation}
which implies $A\propto \mathcal{A}^{-1/2}$. 

The transport equations~\eqref{trans:Phi} and~\eqref{trans:A} provide a way to 
construct a wave solution from a single ray; therefore, any solution to the 
wave equation~\eqref{eq:wave} in a 4D spacetime region can be found from a
three-parameter family of null rays (with smoothly varying initial positions 
and wave vectors) by assigning smoothly varying initial values of $(\Phi,A)$
and then transporting these values along the rays. 
(We use the phrase ``smoothly varying'' to mean that the values of $(\Phi,A)$
must change on the scale of $\mathcal{L} \gg \lambdabar$.)
We note it is often convenient to divide the three-parameter family of initial 
positions of the null rays into two-parameter families of rays with constant 
initial values of $\Phi$.
The constant-$\Phi$ surfaces are the initial phase fronts, which, upon 
propagation along the rays, become 3D phase fronts of the globally defined 
wave. 
The more usual 2D phase fronts, at a given time, are obtained if we take a 
particular time slicing of the spacetime and find the 2D cross sections of the
3D phase fronts in this slicing. 

The above formalism describes wave propagation up to next-to-leading order in 
$\mathcal L/\lambdabar$, which will be enough for us to build a geometric 
correspondence for both the real frequency, the decay rate, and the angular 
separation constant of QNMs in the Kerr spacetime.

\subsection{Null geodesics in the Kerr spacetime}

Now let us review the description of null geodesics in the Kerr spacetime using
the Hamilton-Jacobi formalism. 
In general, the Hamilton-Jacobi equation states
\begin{equation}
\label{HJ}
g^{\mu\nu}(\partial_\mu S)(\partial_\nu S)=0\,, 
\end{equation}
where $S(x^\mu)$ is called the {\it principal function}. 
For the Kerr spacetime, the Hamilton-Jacobi equation can be solved via 
separation of variables (see, e.g., \cite{Chan}), through which the principal
function can be expressed as
\begin{equation}
S(t,\theta, \phi, r) = S_{\theta}(\theta) +L_z \phi+S_r(r)-\mathcal{E} t\,, 
\label{eq:HJFunction}
\end{equation}
where $\mathcal{E}$ and $L_z$ are constants that are conserved because of the
the timelike and axial Killing vectors of the Kerr spacetime.
Physically, $\mathcal E$ and $L_z$ represent the energy and $z$-directed 
specific angular momentum of the massless scalar particle. 
The functions $S_r(r)$ and $S_\theta (\theta)$ are given by
\begin{subequations}
\begin{align}
\label{eq3}
S_r(r) =&\int^r \frac{\sqrt{\mathcal{R}(r')}}{\Delta(r')}dr', & 
S_{\theta} (\theta)&=\int^\theta \sqrt{\Theta(\theta')}d\theta'\,,
\end{align}
where $\mathcal{R}(r)$ and $\Theta(\theta)$ are given by
\begin{align}
\label{eqtheta}
\mathcal{R} (r)&= [\mathcal{E}(r^2+a^2)-L_z a]^2-\Delta[(L_z-a \mathcal{E})^2
+\mathcal{Q}]\,,  \\
\Theta (\theta) &= \mathcal{Q} -\cos^2{\theta}(L^2_z/\sin^2{\theta}-a^2\mathcal{E}^2)\, ,
\end{align}
\end{subequations}
and $\Delta(r)$ is given in Eq.\ (\ref{eqexplan}).
The constant $\mathcal{Q}$ is the Carter constant of the trajectory, which is 
a third conserved quantity along geodesics in the Kerr spacetime.

The principal function $S(x^\mu;\mathcal{E}, L_z,\mathcal{Q})$ contains 
information about all null geodesics; equations of motion for individual null 
geodesics are given by first choosing a particular set of 
$(\mathcal{E},L_z,\mathcal{Q})$, and then imposing 
\begin{equation}
\frac{\partial S}{\partial{\mathcal{E}}} =0\,,\ 
\frac{\partial S}{\partial{L_z}} =0\,,\
\frac{\partial S}{\partial{\mathcal{Q}}} =0\,.\
\end{equation}
These conditions lead to at set of first-order differential equations
\begin{subequations}
\begin{align}
\frac{d t}{d\lambda} =& 
\frac{r^2+a^2}{\Delta}
\left[\mathcal{E}(r^2+a^2)-L_z a\right]-a(a\mathcal{E}\sin^2\theta -L_z)\,,\label{eq:tOrbit} \\
\frac{d \phi}{d\lambda} =& 
-\left(a\mathcal{E}-\frac{L_z}{\sin^2\theta}\right)
+\frac{a \left[\mathcal{E}(r^2+a^2)-L_z a\right] }{\Delta}
\, ,\label{eq:phiOrbit} \\
\frac{d r}{d\lambda} = &{\sqrt{\mathcal{R}}}, \qquad \qquad
 \frac{d \theta}{d\lambda} = \sqrt{\Theta} \, ,
\label{eq:RThetaOrbit}
\end{align}
\end{subequations}
where we have defined
\begin{equation}
\frac{d}{d\lambda} \equiv \Sigma \frac{d}{d\zeta}\,,\quad 
\Sigma =r^2 +a^2\cos^2\theta \,,
\end{equation}
and $\zeta$ is an affine parameter along the null geodesics.

\subsection{Correspondence with quasinormal modes}

\begin{table*}
\caption{Geometric-optics correspondence between the parameters of a 
quasinormal mode, ($\omega$, $A_{lm}$, $l$, and $m$), and the
conserved quantities along geodesics, ($\mathcal E$, $L_z$, and $\mathcal Q$).
To establish a correspondence with the next-to-leading-order, geometric-optics 
approximation, the geodesic quantities $\mathcal E$ and $\mathcal Q$ must be 
complex.}
\begin{tabular}{l*{8}{c}r}
\toprule
&& Wave Quantity && Ray Quantity && Interpretation\\
\hline
&&  $\omega_R$  && $\mathcal{E}$ && 
\begin{tabular}{c} Wave frequency is same as energy of null ray  \\ 
(determined by spherical photon orbit). \end{tabular}  \\
\hline
&& $m$ && $L_z$ && 
\begin{tabular}{c} 
Azimuthal quantum number corresponds to $z$ angular momentum \\ 
(quantized to get standing wave in $\phi$ direction). \end{tabular} \\
\hline
&& $A_{lm}^R$ && $\mathcal{Q}+L^2_z $ && 
\begin{tabular}{c} 
Real part of angular eigenvalue related to Carter constant \\
(quantized to get standing wave in $\theta$ direction). \end{tabular} \\
\hline \hline 
&& $ \omega_I$ && $\gamma=-\mathcal E_I$ && 
\begin{tabular}{c} Wave decay rate is proportional to Lyapunov exponent \\ 
of rays neighboring the light sphere. \end{tabular} \\
\hline
&& $A_{lm}^I$ && $\mathcal Q_I$ && \begin{tabular}{c} 
Nonzero because $\omega_I \neq 0$  \\ (see Secs.\ \ref{sec:ComplexAlm} and 
\ref{sec:ImAngEigenVal} for further discussion). \end{tabular} \\
\hline
\end{tabular}
\label{tb:GeoMatch}
\end{table*}

Here we will find connection between the general set of wave solutions in the
previous section, and the particular solutions that correspond to a quasinormal
modes, in the geometric-optics limit.
Specifically, we will look for waves that propagate outwards at infinity and
down the horizon.
With this correspondence, we will be able to make a geometric interpretation 
of our WKB results from Sec.~\ref{sec:WKB}.

\subsubsection{Leading order: conserved quantities of rays and the real parts 
of quasinormal-mode parameters}
\label{sec:LeadingOrder}

It is straightforward to note that the Hamilton-Jacobi equation~\eqref{HJ} is 
identical to the leading-order geometric-optics equations, if we identify the 
phase, $\Phi$, with the principal function, $S$. 
Therefore, at leading order, we can write
\begin{equation}
\label{ulead}
u = e^{iS} = e^{-i\mathcal{E} t} e^{i L_z\phi} e^{\pm iS_\theta} 
e^{\pm iS_r} \,,
\end{equation}
where we recall that the amplitude $A$ differs from unity only at 
next-to-leading order (we will treat it in the next subsections). 
Here, we have a chosen set of conserved quantities, 
$(\mathcal{E}, \mathcal{Q}, L_z)$, to identify the wave we wish to connect 
with a quasinormal-mode solution.  
The region in which the wave propagates is identical to the region in which 
geodesics with these conserved quantities can propagate.  
In addition, for each point in this region, there is one (and only one) 
geodesic passing through it; that we have $\pm $ in front of $S_\theta$ and 
$S_r$ means only that either propagation direction could be a solution to the 
wave equation.  

Now we note that $u$, a scalar wave in the Kerr spacetime, must separate as in 
Eq.~\eqref{eq6}. 
By comparing Eq.~\eqref{eq6} and Eq.~\eqref{ulead}, we can immediately identify
that  
\begin{equation}
\mathcal{E} = \omega_R\,.
\end{equation}
Because $\mathcal E$ is a real quantity (the conserved energy of the null 
geodesic), we see that at leading order, the wave does not decay. 
Next, we also observe that in order for $u$ to be consistently defined in the 
azimuthal direction, $L_z$ (of the null geodesics that $S$ describes) must be 
an integer.  This allows us to make the second identification 
\begin{equation}
L_z =  m \,.
\end{equation}
Comparing $S_\theta$ from Eq.~\eqref{eq3} and $u_\theta$ from Eqs.\
\eqref{eq:uthetaWKB} and~\eqref{eq:Vtheta} (focusing on one direction of 
$\theta$ propagation, and ignoring next-to-leading-order terms), we can also 
conclude that 
\begin{equation}
\mathcal{Q} =A_{lm}^R -m^2 \,.
\end{equation}
At this stage, given any set of $(\mathcal{E},\mathcal{Q},L_z)$, we will be 
able to find a wave solution that exists in the region in which the geodesics 
travel.  
Not all such sets of conserved quantities correspond to quasinormal modes, 
however, because they may not satisfy the correct boundary conditions of QNMs.

We will first explain the conditions on the radial motion of the geodesics that
will allow these particular geodesics to correspond to a wave that satisfies 
outgoing and downgoing conditions at $r_*\rightarrow \pm \infty$, respectively.
If the radial geodesics satisfy $\mathcal{R}>0$ everywhere, then there will be 
traveling waves across the entire $r_*$ axis, which will not satisfy the 
boundary conditions; if there are two disconnected regions of traveling waves, 
however, waves will scatter off the potential on each side, and they will also
fail to satisfy the boundary conditions.  
The only way to satisfy the boundary conditions is to have a point $r_0$ at 
which $\mathcal{R}=0$ and $\mathcal{R}'=0$, in which case there will be a
family of geodesics on each side of $r=r_0$ (with each member a 
{\it homoclinic orbit} which has $r\rightarrow r_0$ on one end) and a 
{\it spherical orbit} with constant $r=r_0$. 
The corresponding wave has zero radial spatial frequency at $r=r_0$, and this
frequency increases towards $r<r_0$ and decreases towards $r>r_0$.  
Noting that
\begin{equation}
{\mathcal{R}}  = V^r \left(r^2+a^2\right)^2 \,,
\end{equation}
the condition 
\begin{equation}
\label{cond:R}
\mathcal{R} =\mathcal{R}'=0 
\end{equation}
is the same as the condition, Eq.~\eqref{Vreq}, which determines $\omega_R$ in 
terms of $L$ and $m$ in the WKB approximation.  
It is worth mentioning that although the condition of Eq.~\eqref{cond:R} 
imposed on $(\mathcal{E}, \mathcal{Q}, L_z)$ can be interpreted most easily as 
the condition for a spherical photon orbit, the wave function for the 
quasinormal mode we are considering is {\it not} localized around that orbit.  
The wave function at leading order, in fact, has a constant magnitude at every 
location that homoclinic orbits reach (i.e., the entire $r$ axis). 
We will derive the amplitude corrections in the next section.

The quantization of the frequency $\omega_R$ in terms of the multipolar indices
$l$ and $m$ arises from the quantization of the motion in the angular 
directions.  
For the azimuthal direction, it is easy to see that for the wave function to be
single-valued, we need to impose $L_z = m\in\mathbb{Z}$. 
For the $\theta$ direction, we note that 
\begin{equation}
\Theta = V^\theta \sin^2\theta \,,
\end{equation}
and the $\theta$-quantization condition for the wave, Eq.~\eqref{Aeq}, is
\begin{equation}
\int_{\theta_-}^{\theta_+} \sqrt{\Theta}\, d\theta = (L-|m|)\pi\,.
\label{eq:BSOptics}
\end{equation}
This corresponds to the Bohr-Sommerfeld condition for a particle moving in a
potential given by $\Theta$.  
Consequently, the condition for a standing wave along the $\theta $ direction 
(at leading order) is equivalent to 
\begin{align}
\mathcal{Q} &= A_{lm} (\omega_R a) - m^2 \nonumber\\
& \approx L^2 -m^2 -\frac{a^2\omega_R^2}{2}\left[1-\frac{m^2}{L^2}\right] \,.
\end{align}

In summary, we connected the QNM's wave function to the Hamilton-Jacobi 
principal function of homoclinic null geodesics (at leading order).
These geodesics have the same energy, Carter constant, and $z$-component of its
angular momentum as a spherical photon orbit; however only spherical orbits 
with {\it quantized} Carter constants and $z$-angular momenta correspond to 
quasinormal modes. 
In Table~\ref{tb:GeoMatch}, we summarize our geometric-optics correspondence; 
so far we have identified the first three entries on the table. 
We can find the next two correspondences by investigating next-to-leading-order
geometric optics in the next part.

\subsubsection{Next-to-leading order: radial amplitude corrections and the 
imaginary part of the frequency}

\begin{figure}
\includegraphics[width=0.95\columnwidth]{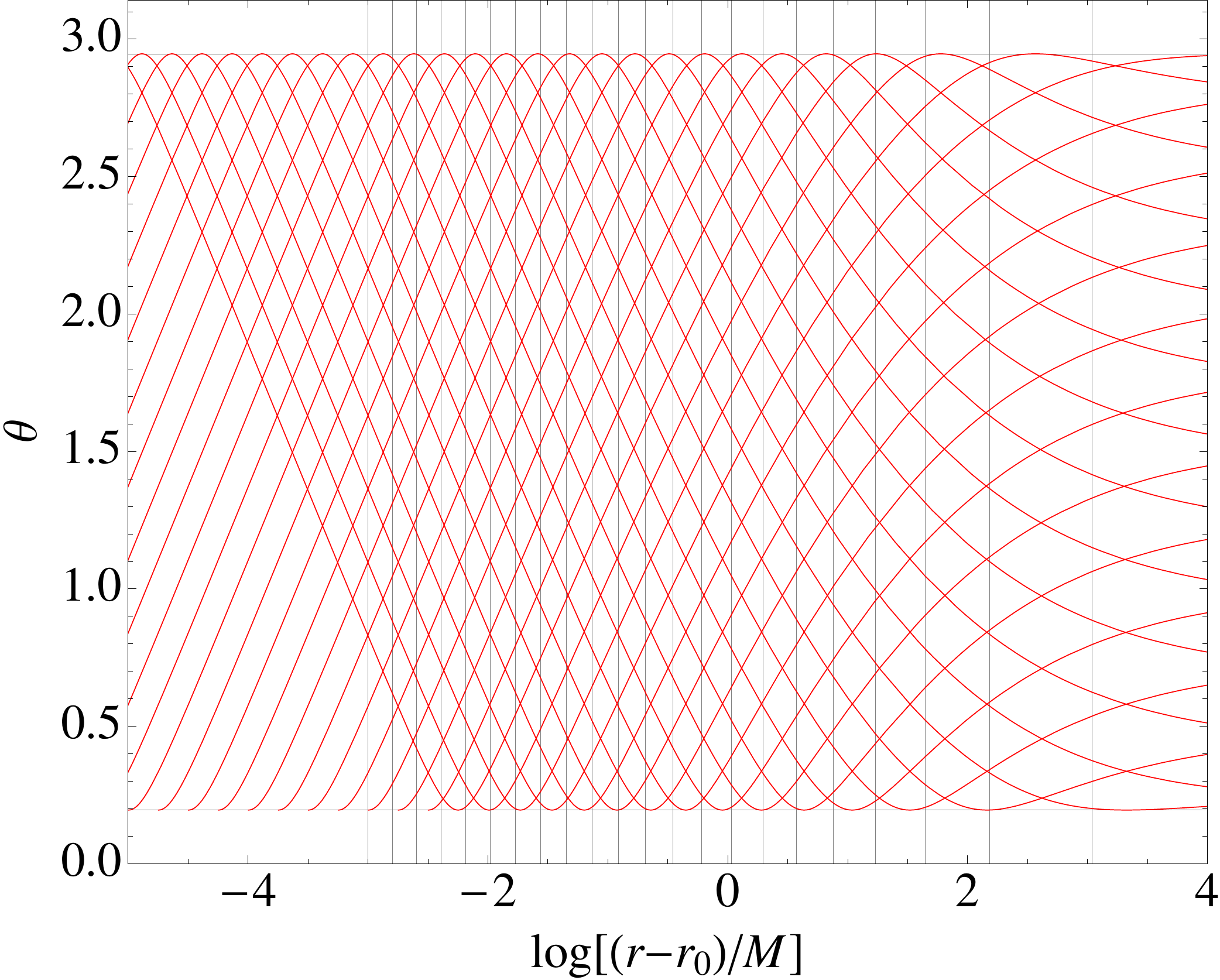}
\caption{Schematic plot of trajectories in the $r$-$\theta$ plane of 
homoclinic orbits outside of the peak of the potential (specifically for a 
black hole with spin $a/M=0.7$ and a photon orbit with radius $r_0/M=2.584$). 
The two horizontal grid lines mark the turning points, $\theta =\theta_\pm$; 
between these turning points, there are two homoclinic orbits passing through
every point, while at turning points only one orbit passes through. 
Vertical grid lines indicate when the value of parameter $\lambda$ has 
changed along the orbit by (an arbitrarily chosen value) 
$\Delta \lambda = 0.046M$.  
Near the spherical photon orbit, each homoclinic orbit undergoes an infinite 
number of periodic oscillations in $\theta$ while $r-r_0$ is growing 
exponentially as a function of $\lambda$.}
\label{homoclinic}
\end{figure}

We showed in the previous part that the conserved quantities of a spherical 
photon orbit, $(\mathcal{E},\mathcal{Q},L_z)$, correspond simply to the real
parts of the quasinormal-mode parameters, $(\omega_R,A_{lm}^R,m)$, which are
the leading-order quantities of a quasinormal mode.
Here, we will show that the behavior of the homoclinic orbits---namely, how 
the orbits propagate away from the spherical orbit, and how they move between 
$\theta_\pm$---reveals the spatiotemporal variation of the wave (i.e, the 
decay rate and the shape of its wave function in space). 
In Fig.~\ref{homoclinic}, we plot the trajectory of a particular series of 
homoclinic orbits on the $r$-$\theta$ plane, to which we will refer at several
points in the discussion below.

With the appropriate values of $(\mathcal{E},\mathcal{Q},L_z)$, the function 
$u$ in Eq.~\eqref{ulead} solves the wave equation to leading order and 
satisfies the required boundary conditions.  
To recover the decaying behavior of quasinormal modes, however, we make 
corrections to the amplitude of the wave, which appear at next-to-leading 
order in the geometric-optics approximation.
Because of symmetry, there should not be any correction to the amplitude in 
the $\phi$ direction, and the correction in the $t$ direction should be a 
simple decay; therefore, we write
\begin{equation}
u =A \exp(iS ) = \underbrace{e^{- \gamma t} A_r(r) A_\theta(\theta)}_{A(t,r,\theta)} e^{-i\mathcal{E} t}  e^{i L_z\phi} e^{\pm iS_\theta} e^{\pm iS_r}\,.
\end{equation}
This general expression contains four possible directions that the wave could 
be propagating: the $\pm\theta$ direction and the $\pm r$ direction (depending 
on the signs in front of $S_r$ and $S_\theta$).  
Because the boundary conditions require that the waves propagate towards 
$r_* \rightarrow +\infty$ for $r>r_0$ and $r_* \rightarrow -\infty$ for 
$r<r_0$, the sign in front of $S_r$ should be positive for $r>r_0$ and negative
for $r<r_0$.  
For $\theta$ motion, however, we insist that both directions (signs) be 
present, because a quasinormal mode is a standing wave in the $\theta$ 
direction.  
Focusing on $r>r_0$, we write 
\begin{align}
u &= e^{-\gamma t} A_r(r) \left[
A^+_\theta e^{iS_\theta} + A^-_\theta e^{-iS_\theta}\right] e^{-i\mathcal{E} t +i L_z \phi + iS_r } \nonumber\\
& \equiv u_+ + u_-\,.
\end{align}

We will next require that both $u_+$ and $u_-$ satisfy the wave equation to 
next-to-leading order, separately. 
By explicitly computing Eq.~\eqref{eik:A} 
(or $A \sqrt{\mathcal{A}} =\mathrm{const}$) in the Kerr spacetime, we find 
the amplitude satisfies the relation
\begin{equation}
\Sigma \frac{d \log A}{d\zeta} =-\frac{1}{2}\left[ \partial_r(\Delta (r)\partial_rS_{r})  
+\frac{1}{\sin\theta}\partial_\theta(\sin\theta \partial_\theta S_\theta)
\right] \,.
\label{eq:AmpChange}
\end{equation}
Here $\zeta$ is an affine parameter along the geodesic specified by 
$(\mathcal{E},\mathcal{Q},\mathcal{L}_{z})$. 
If we use the parameter $\lambda$ defined by $d/d\lambda = \Sigma d/d\zeta$ 
then we can separate the left-hand side of the equation as
\begin{align}
\label{dev}
\Sigma \frac{d \log A}{d\zeta} = \frac{d}{d\lambda} \log A_{r}(r)  + \frac{d}{d\lambda} \log A_{\theta}(\theta) -\gamma \frac{dt}{d\lambda} \,.
\end{align}

Because the right-hand side of Eq.~\eqref{eq:tOrbit} for $dt/d\lambda$, 
separates into a piece that depends only upon $r$ and one that depends only 
upon $\theta$, we will write Eq.\ \eqref{eq:tOrbit} schematically as 
\begin{equation}
\frac{dt}{d\lambda} = \overline{\dot t} +  \tilde{\dot t} \, ,
\label{eq:dtdlambda}
\end{equation}
where $\overline{\dot t}$ is only a function of $r$ and $\tilde{\dot t}$ is 
only a function of $\theta$.
Unlike in Eq.~\eqref{eq:tOrbit}, we will require that $\tilde{\dot t}$ average
to zero when integrating over $\lambda$ for half a period of motion in the 
$\theta$ direction (i.e., from $\theta_-$ to $\theta_+$). 
We can ensure this condition is satisfied by subtracting an appropriate 
constant from $\tilde{\dot t}$ and adding it to $\overline{\dot t}$.  
Combining Eqs.\ (\ref{eq:AmpChange})--(\ref{eq:dtdlambda}) and performing a 
separation of variables, we obtain
\begin{subequations}
\begin{align}
\sqrt{\mathcal{R}}\frac{d\log A_r}{dr}  -\gamma \overline{\dot t}  &=-\frac{\mathcal{R}'}{4\sqrt{\mathcal{R}}}\,,
\label{eq:RAmp} \\
 \sqrt{\Theta} \frac{d \log A_\theta^\pm}{d\theta} \mp \gamma\tilde{\dot t} &= 
-\frac{1}{2\sin\theta}( \sqrt{\Theta} \sin\theta)'  \,,
\label{eq:ThetaAmp}
\end{align}
\end{subequations}
where a prime denotes a derivative with respect to $r$ for functions of $r$ 
only, and a derivative with respect to $\theta$ for functions of $\theta$ only
(whether it is a $\theta$ or $r$ derivative should be clear from the context). 
While it might at first seem possible to add a constant to the definition of 
$\overline{\dot t}$, and subtract it from $\tilde{\dot t}$ and still have both 
$u_+$ and $u_-$ satisfy the next-to-leading order geometric optics, because we
have already chosen to have $\tilde {\dot t}$ average to zero,
\begin{equation}
\int_{\theta_-}^{\theta_+}\gamma\tilde{\dot t}\frac{d\theta}{\sqrt{\Theta}} = \int \gamma \tilde{\dot t} d\lambda =0 \,,
\end{equation}
this separation is the only way to guarantee that $|A_\theta^\pm|$ match each 
other at both ends.  
We will discuss the angular wave function in greater detail in the next part of
this section.

Let us now turn to the radial equation, from which we will be able to compute
the decay rate.
Close to $r_0$, we can expand $\mathcal R(r)$ to leading order as 
\begin{equation}
\mathcal{R}(r) \approx \frac{(r-r_0)^2}{2}\mathcal{R}''(r_0) \, .
\label{eq:RquadExp}
\end{equation}
Substituting this result into Eq.~\eqref{eq:RAmp}, we find
\begin{equation}
\frac{d\log A_r}{dr} =\frac{1}{r-r_0}\left[\gamma \overline{\dot t} \sqrt{\frac{2}{{\mathcal{R}''_0}}} -\frac{1}{2}\right]\,,
\end{equation}
where we used the notation $\mathcal{R}''_0 \equiv \mathcal{R}''(r_0)$.  
For $A_r$ to be a function that scales as $A_r \sim (r-r_0)^n$ around $r_0$ 
for some integer $n$ (namely it scales like a well-behaved function), we need 
to have
\begin{align}
\label{gammanew}
\gamma &=\left(n+\frac{1}{2}\right) \frac{\sqrt{\mathcal{R}''_0/2} }{\overline{\dot t}} \nonumber\\
&=(n+1/2)\lim_{r\rightarrow r_0} \frac{1}{r-r_0} \frac{dr/d\lambda}{\langle dt/d\lambda \rangle_\theta} \,.
\end{align} 
To arrive at the second line, we used Eq.\ (\ref{eq:RquadExp}), the fact that
$dr/d\lambda = \sqrt{\mathcal R}$, and that $\overline{\dot t}$ is the part of
$dt/d\lambda$ that does not vanish when averaging over one cycle of motion in
the $\theta$ direction; the limit in the expression comes from the fact that
the approximation in Eq.\ (\ref{eq:RquadExp}) becomes more accurate as 
$r\rightarrow r_0$.

The physical interpretation of the rate that multiplies $(n+1/2)$ is somewhat
subtle.
Because the $\theta$ motion is independent from $r$ motion, a bundle of 
geodesics at the same $r$ slightly larger than $r_0$, but at different 
locations in $\theta$, will return to their respective initial values of 
$\theta$ with a slightly increased value of $r$ after one period of motion in
the $\theta$ direction.
The area of this bundle increases in the process, and by Eq.\ \ref{trans:A}, 
the amplitude of the wave must decay; the rate of decay is governed by the
quantity that multiplies $(n+1/2)$ in Eq.\ (\ref{gammanew}).
  
In addition, as shown in Fig.~\ref{homoclinic}, the homoclinic orbits do pass
through an infinite number of such oscillations near $r_0$, because the radial 
motion is indefinitely slower than the $\theta$ motion as $r$ approaches $r_0$.
It is clear from Fig.~\ref{homoclinic} that
\begin{equation}
\frac{1}{r-r_0}\frac{\Delta r}{\Delta \lambda} = \frac{\Delta \log (r-r_0)}{\Delta \lambda}
\end{equation}
approaches a constant as $r\rightarrow r_0$.
By multiplying the above equation by the constant value of
$(\Delta \lambda)/(\Delta t)$ over one orbit of motion in the $\theta$ 
direction,
\begin{equation}
\frac{1}{r-r_0}\frac{\Delta r}{\Delta t} = \frac{\Delta \log (r-r_0)}{\Delta t} \equiv \gamma_{L}
\label{eq:Lyapunov}
\end{equation}
also approaches a constant.  
This is usually defined as the {\it Lyapunov exponent} of one-dimensional 
motion; here, however, we emphasize that it is defined only after averaging 
over entire cycle of $\theta$ motion. 
By comparing Eq.\ (\ref{eq:Lyapunov}) with the second line of Eq.\ 
(\ref{gammanew}), and bearing in mind that the Lyapunov exponent is defined
after averaging over one period of $\theta$ motion, one can write Eq.\ 
(\ref{gammanew}) as
\begin{equation}
\gamma = (n + \tfrac 12) \gamma_L \, .
\label{eq:gammaLyapunov}
\end{equation}

To put Eq.~\eqref{gammanew} into a form that relates more clearly to 
Eq.~\eqref{omegai}, we use the conditions on the phase function,
\begin{equation}
\frac{\partial S}{\partial \mathcal{E}}=0\,,\quad 
\frac{\partial S}{\partial \mathcal{Q}}=0\,,
\label{eq:dSdEdSdQ}
\end{equation}
which hold for any point on the trajectory of the particle.  
We will apply this condition to two points on the particle's trajectory: one 
at $(t,r,\theta,\phi)$ and the second at 
$(t+\Delta t,r+\Delta r, \theta,\phi+\Delta\phi)$, where $\Delta t$ is chosen 
such that the particle completes a cycle in $\theta$ in this time (and it moves
to a new location shifted $\Delta r$ and $\Delta\phi$).  
Substituting in the explicit expressions for the principal function in Eqs.\
(\ref{eq:HJFunction}) and (\ref{eq3}), we find
\begin{subequations}
\begin{align}
\label{eqtE}
\frac{\partial}{\partial \mathcal{E} }
\left[ 
\int_{r}^{r+\Delta r}\frac{\sqrt{\mathcal{R}(r')}}{\Delta(r')} dr' 
+ \Delta S_\theta \right]   =\Delta t\, \\
\label{eqtQ}
\frac{\partial}{\partial \mathcal{Q} }
\left[ 
\int_{r}^{r+\Delta r}\frac{\sqrt{\mathcal{R}(r')}}{\Delta(r')} dr' 
+ \Delta S_\theta \right]   =0\, \,.
\end{align}
\end{subequations}
where we have defined
\begin{equation}
\Delta S_\theta \equiv 2\int_{\theta_-}^{\theta_+}  \sqrt{\Theta(\theta')} 
d\theta' \, \equiv \oint \sqrt{\Theta(\theta')} d\theta'.
\end{equation}
Because the change $\Delta r$ is infinitesimal for $r$ near $r_0$, the 
integrand is roughly constant, and the $r$-dependent part of the integral 
becomes the product of the integrand with $\Delta r$.
Then, one can use Eq.\ (\ref{eq:RquadExp}) to write Eqs.\ (\ref{eqtE}) and 
(\ref{eqtQ}) as
\begin{subequations}
\begin{align}
\label{eqt1}
\frac{1}{\sqrt{2\mathcal{R}''_0} \Delta_0 }\frac{\partial\mathcal{R}}{\partial\mathcal{E}}\frac{\Delta r}{r-r_0} + \frac{\partial \Delta S_\theta}{\partial \mathcal{E}}=\Delta t \,, \\
\label{eqt2}
\frac{1}{\sqrt{2\mathcal{R}''_0} \Delta_0 }\frac{\partial\mathcal{R}}{\partial\mathcal{Q}}\frac{\Delta r}{r-r_0} + \frac{\partial \Delta S_\theta}{\partial \mathcal{Q}}=0\,.
\end{align}
\end{subequations}
Now, we also note that for a given fixed $L_z=m$, the angular Bohr-Sommerfeld 
condition in Eq.\ (\ref{eq:BSOptics}) makes $\mathcal{Q}$ a function of 
$\mathcal{E}$ through the condition that $\Delta S_\theta = (L-|m|)\pi$.
Because $\Delta S_\theta$ is a function of $\mathcal E$, its total derivative
with respect to $\mathcal E$ must vanish,
\begin{equation}
\frac{\partial \Delta S_\theta}{\partial \mathcal{E} } +
\frac{\partial \Delta S_\theta}{\partial \mathcal{Q} } \left(\frac{d\mathcal{Q}}{d\mathcal{E}} \right)_{\rm BS} =0\,.
\label{eq:BSdQdE}
\end{equation}
Therefore, when we multiply Eq.\ \eqref{eqt2} by 
$(d\mathcal Q/d\mathcal E)_{\rm BS}$ and add it to Eq.\ \eqref{eqt1}, we obtain
the condition that
\begin{equation}
\frac{1}{\sqrt{2\mathcal{R}''_0} \Delta_0 }
\left[\frac{\partial\mathcal{R}}{\partial\mathcal{E}} +
\frac{\partial\mathcal{R}}{\partial\mathcal{Q}} \left(\frac{d\mathcal{Q}}{d\mathcal{E}} \right)_{\rm BS}
\right]
\frac{\Delta r}{r-r_0} =\Delta t \,.
\end{equation}
Combining this fact with the definition of the Lyapunov exponent in Eq.\ 
(\ref{eq:Lyapunov}) and Eq.\ (\ref{eq:gammaLyapunov}), we find that
\begin{equation}
\label{gammawkb}
\gamma =\left(n+\frac{1}{2}\right)  \frac{\sqrt{2 \mathcal{R}''_0}\Delta_0 }{\displaystyle\left[\frac{\partial\mathcal{R}}{\partial\mathcal{E}} +
\frac{ \partial\mathcal{R}}{\partial\mathcal{Q}} \left(\frac{d\mathcal{Q}}{d\mathcal{E}}\right)_{\rm BS}\right]_{r_0}} \,,
\end{equation}
where we recall that the quantities should be evaluated at $r_0$.
Equation\ (\ref{gammawkb}) is equivalent to Eq.~\eqref{omegai}.
Note, however, that in Eq.\ (\ref{gammawkb}) we explicitly highlight the 
dependence of $\mathcal Q$ on $\mathcal E$ through the term 
$(d\mathcal{Q}/d\mathcal{E})_{\rm BS}$.
There is an analogous term in Eq.~\eqref{omegai} from the dependence of 
$A_{lm}$ on $\omega$ in the expression for the potential $V^r$, which we must 
take into account when computing $\partial V^r/\partial \omega$; however, we
did not write it out explicitly in Eq.~\eqref{omegai}.

Summarizing the physical interpretation of the results in this subsection, we 
note that the Lyapunov exponent $\gamma_L$ is the rate at which the 
cross-sectional area of a bundle of homoclinic rays expand, when averaged over 
one period of motion in the $\theta$ direction in the vicinity of $r_0$.
The spatial Killing symmetry along $\phi$ means the extension of the ray bundle
remains the same along that direction.  
This, therefore, allows us to write
\begin{equation}
\mathcal{A} \sim e^{\gamma_L t} \,.
\end{equation}
Correspondingly, the $A\sqrt{\mathcal{A}}=\mathrm{const}$ law requires that 
\begin{equation}
A \sim e^{-\gamma_Lt/2} \,,
\end{equation}
which agrees with the decay rate of the least-damped QNM.  
The higher decay rates given by Eq.~\eqref{gammanew} come from an effect 
related to the intrinsic expansion of the area of a phase front.
More specifically, if the amplitude is already nonuniform at points with 
different $r-r_0$ (but same $\theta$), then shifting the spatial locations of 
the nonuniform distribution gives the appearance of additional decay.

\subsubsection{Next-to-leading order: angular amplitude corrections and the 
imaginary part of Carter's constant}
\label{sec:ImAngEigenVal}

Having found a relation in Eq.~\eqref{eq:RAmp} between the imaginary part of 
the energy, $\omega_I$, and the rate of divergence of rays, we now turn to 
Eq.~\eqref{eq:ThetaAmp} to understand the geometric meaning of the complex part
of $A_{lm}$.  
We recall from Sec.~\ref{sec:LeadingOrder} that $\mathcal{Q} =A_{lm}^R-m^2$, at
leading order, for a real Carter constant $\mathcal Q$.
Because $A_{lm}$ becomes complex at next to leading order (and because $m$
remains unchanged), if the correspondence $\mathcal Q=A_{lm}-m^2$ holds for a 
complex $A_{lm}$, then the Carter constant should also be complex, and its 
imaginary part should be equivalent to $A_{lm}^I$.
In this part, we argue that this relationship holds.

By integrating Eq.~\eqref{eq:ThetaAmp}, we find that
\begin{equation}
A_\theta^\pm = \sqrt{\frac{1}{\sin\theta\sqrt{\Theta}}} \exp\left[\pm \int^\theta_{\theta_-}  \frac{\gamma \tilde{\dot t}}{\sqrt{\Theta}}d\theta'\right] \,.
\label{eq:ApmTheta}
\end{equation}
To interpret this equation, we will assume that the orbit is sufficiently close
to $r_0$ that the change in $r$ over the course of a period of motion in 
$\theta$ is negligible.
Under this assumption (and with the fact that $d\lambda = d\theta/\sqrt\Theta$)
we can write the integral in the exponent in Eq.\ (\ref{eq:ApmTheta}) as
\begin{align}
&\int_{\theta_-}^\theta \frac{\gamma \tilde{\dot t}}{\sqrt{\Theta}}d\theta' \nonumber\\
=&
\gamma \left[[t(\theta)-t(\theta_-)] -\left(\frac{\Delta t}{\Delta \lambda}\right) [\lambda(\theta)-\lambda(\theta_-)]\right] \,,
\label{eq:IntTildeDotT}
\end{align}
where $\Delta t / \Delta \lambda$ is the average of $dt/d\lambda$ over a cycle of $\theta$ motion.
We obtain this expression by using the fact that $dt/d\lambda$ is equivalent to 
$\tilde{\dot t}$ plus a constant when $r$ (and hence 
$\overline{\dot t}$) does not change.
Because $\tilde{\dot t}$ has zero average (by definition) over a period of 
$\theta$ motion, then when written in the form above, the constant must be 
$(\Delta t)/(\Delta \lambda)$.
We can write this average rate of change in a useful form by noting that, from 
Eq.~\eqref{eq:tOrbit} and Eqs.~\eqref{eqtheta}, 
\begin{align}
\label{eq:TimeRateSplit}
\frac{dt}{d\lambda} & = \frac{1}{2\Delta} \frac{\partial \mathcal R} {\partial \mathcal E} + a^2 \mathcal E \cos^2 \theta \,.
\end{align}
Averaging this expression over a cycle of $\theta$ motion, noting that the 
first term on the right-hand side is independent of $\theta$, and using 
Eq.~\eqref{eqtheta} gives
\begin{align}
\label{eq:BSDeltaT}
\frac{\Delta t}{\Delta \lambda} & =  \frac{1}{2\Delta} \frac{\partial \mathcal R} {\partial \mathcal E} + a^2 \mathcal E \left( \int_{\theta_-}^{\theta^+} \frac{\cos^2 \theta}{\sqrt{\Theta}}d\theta \right)\left(\int_{\theta_-}^{\theta^+} \frac{d\theta}{\sqrt{\Theta}}\right)^{-1} \nonumber \\
& =  \frac{1}{2\Delta} \frac{\partial \mathcal R} {\partial \mathcal E}  + \frac{\partial \Delta S_\theta / \partial \mathcal E}{2 \partial \Delta S_\theta/\partial \mathcal Q}  \nonumber \\
&= \frac{1}{2\Delta} \frac{\partial \mathcal R} {\partial \mathcal E} - \frac12 \left( \frac{d \mathcal Q}{d\mathcal E} \right)_{\rm BS} \,.
\end{align}
In the last step we have used the Bohr-Sommerfeld condition \eqref{eq:BSdQdE}.
Also according to Eq.~(\ref{eq:tOrbit}) and Eq.~(\ref{eq:RThetaOrbit}), we can 
find 
\begin{subequations}
\begin{align}
t(\theta)-t(\theta_-) =& \frac{\partial }{\partial\mathcal{E}}\int_{\theta_-}^\theta \sqrt{\Theta(\theta')}d\theta'
\nonumber \\ &
+\frac{1}{2\Delta}\frac{\partial \mathcal{R}}{\partial \mathcal{E}}(\lambda(\theta)-\lambda(\theta_-))\label{eq:DeltaT} \,,\\
\lambda(\theta)-\lambda(\theta_-) =&2 \frac{\partial }{\partial\mathcal{Q}}\int_{\theta_-}^\theta \sqrt{\Theta(\theta')}d\theta' \,,
\label{eq:DeltaLambda}
\end{align}
\end{subequations}
where to derive these two equations, we can again use the fact that 
$d\lambda = d\theta/\sqrt{\Theta}$ and the definition of $\Theta$; for the 
first we also make use of Eq.~\eqref{eq:TimeRateSplit}.

Finally, we insert Eqs.~\eqref{eq:DeltaT}, \eqref{eq:DeltaLambda}, and 
\eqref{eq:BSDeltaT} into Eq.~\eqref{eq:IntTildeDotT} to find
\begin{equation}
\int_{\theta-}^\theta \frac{\gamma \tilde{\dot t}}{\sqrt{\Theta}}d\theta'
=(-i \gamma)\left[\frac{\partial }{\partial\mathcal{E}} + 
\left(\frac{d\mathcal{Q}}{d\mathcal{E}}\right)_{\rm BS}\frac{\partial }{\partial\mathcal{Q}}\right]
\left[i S_\theta(\theta)\right]\,.
\label{eq:AmpThetaFinal}
\end{equation}
Substituting Eq.\ (\ref{eq:AmpThetaFinal}) into the solution for $A^\pm_\theta$
in Eq.\ (\ref{eq:ApmTheta}) gives that
\begin{equation}
A_\theta^\pm =  \frac{\exp\left\{ (\pm i \gamma)
\left[\frac{\partial }{\partial\mathcal{E}} + 
\left(\frac{d\mathcal{Q}}{d\mathcal{E}}\right)_{\rm BS}
\frac{\partial }{\partial\mathcal{Q}}\right]
\left[i S_\theta(\theta)\right] \right\}}{\sqrt{\sin \theta \sqrt{\Theta}}}\, .
\label{eq:ApmFinal}
\end{equation}
The phase in this equation, however, is precisely the correction to the 
leading-order expression for the phase $e^{iS_\theta(\theta)}$ if we allow
$\mathcal E$ and $\mathcal Q$ to be complex, where their imaginary parts are 
given by
\begin{equation}
\mathrm{Im} \mathcal{E} = -\gamma =-\omega_I
\,,\quad\mathrm{Im} \mathcal{Q} =\left(\frac{d\mathcal{Q}}{d\mathcal{E}}\right)_{\rm BS} (-\gamma)\,.
\label{eq:ComplexQE}
\end{equation}
Through next-to-leading order, therefore, the $\theta$ portion of the wave is 
given by
\begin{equation}
A_\theta^+ e^{i S_\theta(\theta)} + A_\theta^- e^{-iS_\theta(\theta)} 
= \frac{e^{i S_\theta(\theta)} + e^{-i S_\theta(\theta)}}
{\sqrt{\sin \theta \sqrt{\Theta}}}  \,,
\end{equation}
where $\mathcal{E}$ and $\mathcal{Q}$ used in $S_\theta$ are complex.

In the geometric-optics approximation, therefore, we have shown that we can 
account for the amplitude corrections to the wave by allowing the conserved 
quantities, $\mathcal E$ and $\mathcal Q$, to be complex [with their imaginary
parts given in Eq.\ (\ref{eq:ComplexQE})].
Furthermore, through the geometric-optics correspondence, and the definition
of $A_{lm}^I$ in Eq.\ (\ref{almI}) we can confirm that 
$A_{lm}^I = \mathcal Q_I$; therefore, the relationship
\begin{equation}
\mathcal Q = A_{lm} - m^2 \, ,
\end{equation}
is true for a complex $\mathcal Q$ and $A_{lm}$.

In closing, we note that at the same $\theta$, the magnitude of the two 
components of the wave in Eq.\ (\ref{eq:ApmFinal}) are not equal.  
More specifically, the integral involving $\tilde{\dot t}$ makes $A^+$ have a 
larger amplitude at $\theta <\pi/2$ and a smaller amplitude at $\theta>\pi/2$; 
$A^-$ has the opposite profile.
Therefore, the net wave function remains symmetric about $\theta=\pi/2$.

\section{Features of the Spectra of Kerr Black Holes}
\label{sec:Spectra}

In this section, we will use the WKB formula and the geometric-optics 
correspondence in the first two sections of this paper to explain several
aspects of the quasinormal-mode spectrum of Kerr black holes.
Specifically, we will explain the absence of damping for a significant fraction
of modes of extremal Kerr holes. 
We will also decompose the frequency into orbital and precessional parts and 
explain a degeneracy in the spectra of Kerr holes in terms of a rational 
relation of these frequencies when the corresponding photon orbits close.

\subsection{Spherical photon orbits and extremal Kerr black holes}

We will first review the properties of spherical photon orbits.
These orbits can be found by setting $\mathcal{R}(r)=\mathcal{R}'(r)=0$
(see, e.g., \cite{Chan}), and their conserved quantities are fixed by the 
radius of the orbit $r$ and the spin of the black hole $a$ to be
\begin{subequations}
\label{epsiq}
\begin{align}
\mathcal{Q}/\mathcal{E}^2 & = -\frac{r^3(r^3-6Mr^2+9M^2r-4a^2M)}{a^2(r-M)^2} \
\,, \\
L_z/\mathcal{E} & = -\frac{r^3-3Mr^2+a^2r+a^2M}{a(r-M)} \, .
\end{align}
\end{subequations}
We will next discuss additional features of these orbits.

For a given spin parameter $a$, there is a unique spherical photon orbit with 
parameters $(\mathcal E, L_z, \mathcal Q)$ for any radius between the outermost
and innermost photon orbits (the retrograde and prograde equatorial photon 
orbits).
Their radii (which we denote $r_1$ for retrograde and $r_2$ for prograde 
orbits) are given by 
\begin{subequations}
\begin{align}
r_1 &= 2M \left[ 1+ \cos\left( \frac{2}{3}\arccos\left(-\frac{|a|}{M}\right)
\right) \right] \,,  \\
r_2 &= 2M \left[ 1+ \cos\left( \frac{2}{3}\arccos\left(\frac{|a|}{M}\right)
\right) \right] \,.
\label{eq:EquitorialOrbits}
\end{align}
\end{subequations}
At each $r_1 \le r \le r_2$, the spherical orbit's inclination angle reaches a
maximum and minimum of $\theta_\pm$ (at which $\Theta=0$).
These angles are given by
\begin{align}
 \cos^2 & \theta_{\pm}  = \notag \\
& \frac{\left[2\sqrt{M \Delta(2r^3-3 M r^2+M a^2)}-(r^3-3M^2 r+2M a^2)\right] r}{a^2(r-M)^2} \, ,
\end{align}
which are equivalent to the turning points of the integral~\eqref{Aeq} (and, 
therefore, we use the same symbols for these angles). 

Using the geometric-optics correspondence between $(\mathcal E,L_z,\mathcal Q)$
and $(\Omega_R, \mu, \alpha_{lm}^R)$, we see that equatorial orbits at $r_1$ 
and $r_2$ corresponds to modes with $\mu  =-1$ and $+1$, respectively, or modes
with $m=\pm l$ and $l\gg 1$ (strictly speaking, though, $\mu = m /(l+1/2)$ 
never precisely reaches $\pm 1$). 
We can also relate $r_p$, the real root of Eq.~\eqref{r0}, to the polar orbit 
and modes with $m=0$.  
For orbits between the equatorial and polar ones, we can use Eqs.~\eqref{v} and
\eqref{dvdr} to obtain a $\mu$ between $-1$ and $+1$.
Then, only those modes that can be written as $m/(l+1/2)$ with the allowed 
integer values of $l$ and $m$ correspond to a QNM (though the photon orbits 
that correspond to QNMs are a dense subset of all photon orbits). 

\begin{figure}
\includegraphics[width=0.95\columnwidth]{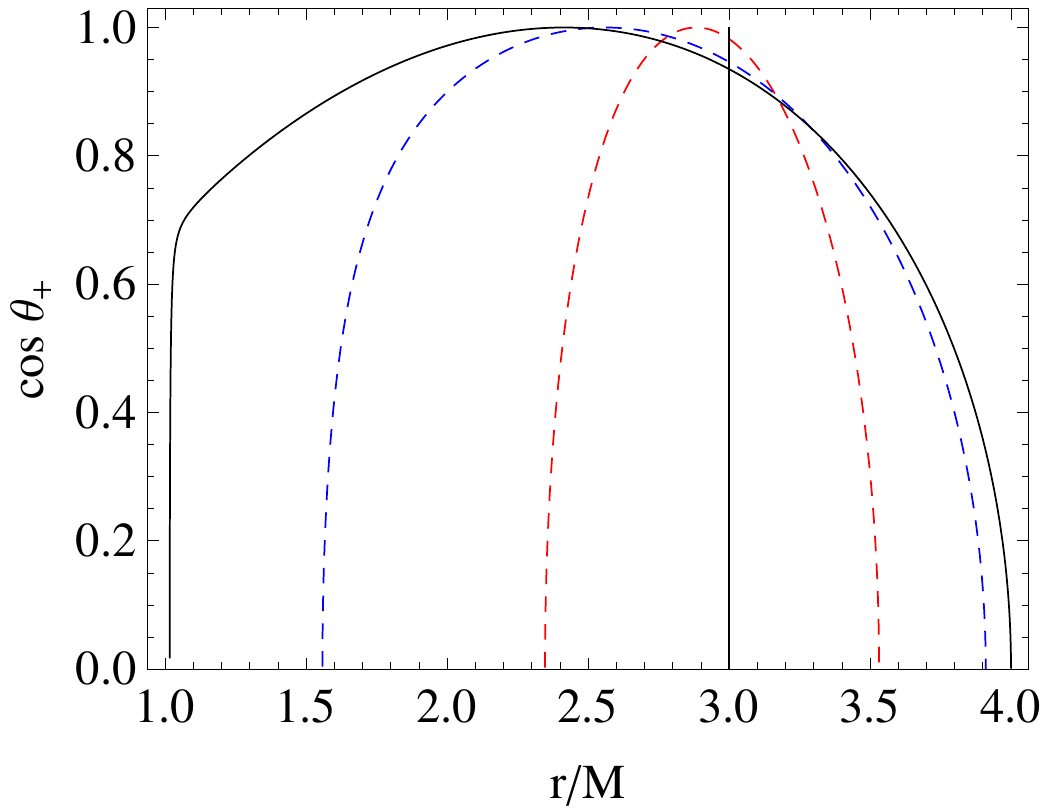}
\caption{The values of $r$ and $\cos\theta_+$ of spherical spherical orbits, 
for $a/M=0$ (black, solid vertical line), $0.5$ [red (light gray) dashed curve],
$0.9$ [blue (dark gray) dashed curve] and $0.99999$ (black, solid curve).  
Note that for $a=0$, all such orbits have $r=3 M$, while for $a=M$, a 
significant fraction reside at $r=M$.}
\label{fig:spheres}
\end{figure}

\begin{figure}
\includegraphics[width=0.95\columnwidth]{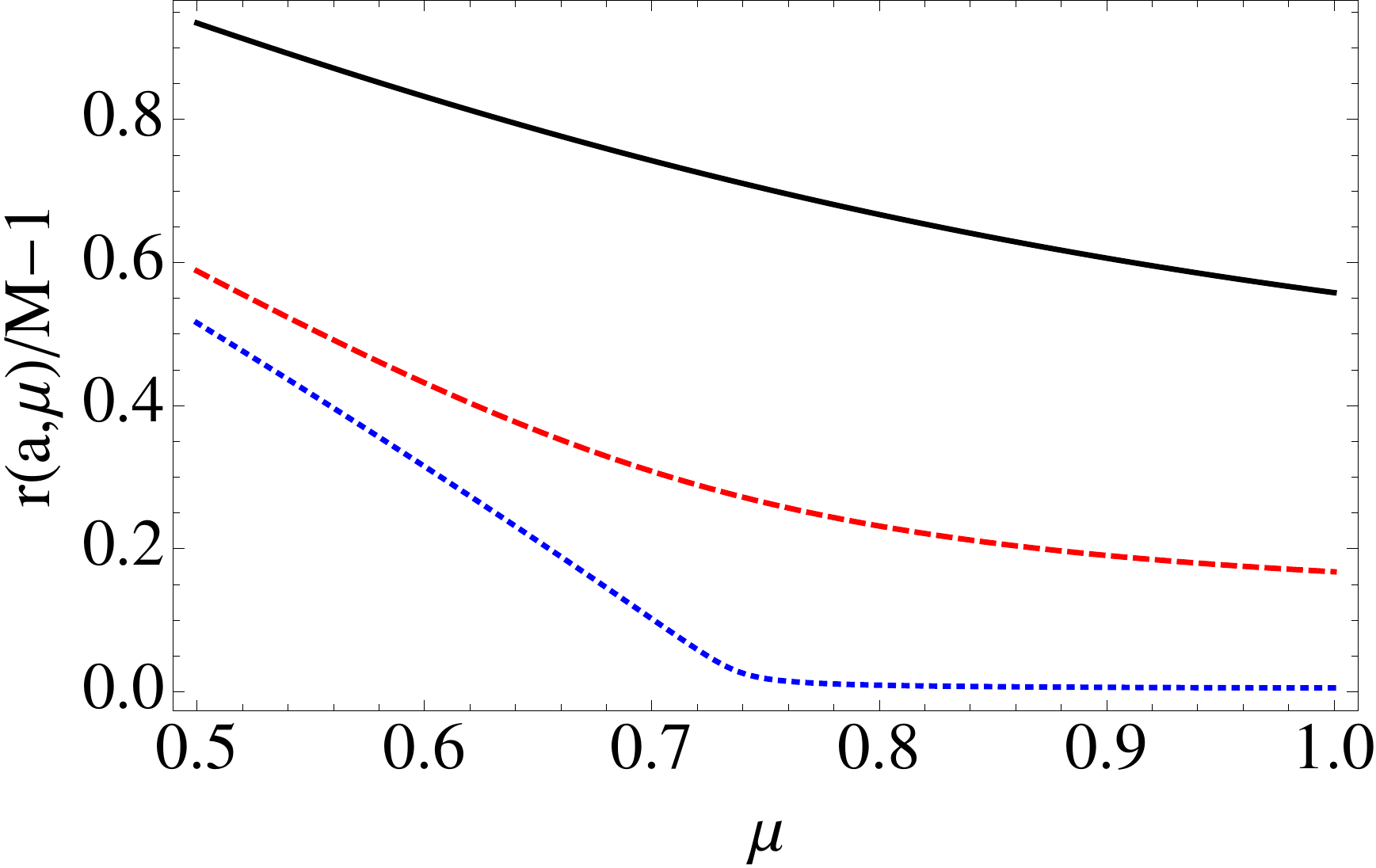}
\caption{Radii of corotating spherical photon orbits as a function of $\mu$, 
for $a/M=0.9$ (black solid line), $0.99$ (red dashed curve), $0.9999$ (blue
dotted line). 
For extremal Kerr black holes, a nonzero fraction of all spherical photon 
orbits are on the horizon.}
\label{fig:radius}
\end{figure}

Note in Fig.\ \ref{fig:spheres} that for $a \sim M$, a significant fraction of
spherical photon orbits of different inclination angles all have nearly the 
same radius, $r \approx M$.  
Through the geometric-optics correspondence, a large fraction of modes (a 
finite range of values of $\mu$) relate to this set of modes with 
$r \approx M$.
In Fig.~\ref{fig:radius}, we explicitly show the relation between modes 
characterized by $\mu$ and their corresponding spherical-photon-orbit radii 
(normalized by the horizon radius) for several values of $a/M$ slightly less 
than unity.
The radius exhibits an interesting transition between two kinds of behaviors:
for $\mu > \mu_* \approx 0.744$, the value of $r$ is very close to $M$ (the 
horizon radius for an extremal Kerr black hole), and for $\mu < \mu_*$ the
radii increase linearly.  
The orbits with $\mu > \mu_*$ have a range of inclination angles.
Their $\sin\theta_\pm$ span from 0.731 (at $\mu_*$, the most inclined orbit) 
to $1$ (at $\mu =1$, the prograde equatorial orbit).  

For the extremal black holes, therefore, a nonzero fraction of corotating 
spherical photon orbits appear to coincide with the horizon in the 
Boyer-Lindquist coordinate system.  
Although the proper distance between these orbits will not vanish (see 
\cite{BardeenPressTeukolsky}), this does not seem to be a coordinate effect,
because there is a definite physical change of the modes for these values of 
$\mu > \mu_*$.
By comparing with Fig.\ \ref{fig:radius} with Fig.\ \ref{fig:OmegaI}, we see 
that these orbits also have $\Omega_I\approx 0$.  
A vanishing imaginary part of the frequency corresponds to a vanishing of the
radial Lyapunov exponent for this entire nonzero region of spherical photon 
orbits.
This, therefore, would lead to a curious effect for a highly spinning black 
hole: for perturbations with $\mu \ge \mu_*$, modes do not move away from or 
into the horizon very quickly. 
If we were to solve an initial-data problem containing these modes, we would
find that they live for a long time; moreover, because these are nearly 
equatorial modes, we would see that the final, long-lived perturbations escape 
in the equatorial direction.  
This would imply that if we were to drive the black hole appropriately with 
equatorial incoming radiation, we would excite these nearly lossless modes with
high amplitude (like optically exciting a resonant cavity with high finesse).

\subsection{A mode's orbital and precessional frequencies}
\label{subsec:modefreq}

In this part, we will define two frequencies associated with individual 
spherical photon orbits (the orbital and precessional frequencies) and 
understand their connection to the real part of the QNM frequency.  
We begin by noting that because spherical photon orbits have only two 
independent degrees of freedom describing their motion [see, e.g., 
Eq.\ (\ref{epsiq})], the orbit can be characterized by two frequencies.  
The first is the $\theta$-frequency, $\Omega_\theta$, the frequency at which 
the particle oscillates below and above the equatorial plane.  
During each $\theta$-cycle, which takes an amount of time given by 
$T_\theta =2\pi/\Omega_\theta$, the particle also moves in the azimuthal 
(or $\phi$) direction. 
If this angle is $2\pi$ for a corotating orbit ($m>0$) or $-2\pi$ for a 
counterrotating orbit ($m<0$), then there is no precession (and these simple,
closed orbits have effectively one frequency describing their motion, as the
spherical photon orbits of a Schwarzschild black hole do).
The difference between the $\Delta \phi$ and $\pm 2\pi$ (its precession-free 
value) we will denote as the {\it precession angle}, 
\begin{equation}
\label{phiprec}
\Delta\phi_{\rm prec} \equiv \Delta\phi - 2 \pi \, \mathrm{sgn} \, m \,,
\end{equation}
where $\mathrm{sgn}\, m$ is the sign of $m$.  
We can also associate the rate of change of $\phi_{\rm prec}$ with a frequency,
\begin{equation}
\Omega_{\rm prec} \equiv \Delta\phi_{\rm prec}/T_\theta = 
\Delta \phi_{\rm prec}\Omega_\theta /(2\pi) \,.
\end{equation} 
Both $T_\theta$ and $\Delta \phi_{\rm prec}$ can be computed from geodesic 
motion [see the formulas for $\Omega_\theta$ and $\Delta\phi_{\rm prec}$ in 
Eq.~(\ref{eqfreqs})]. 

It is possible to perform split of the real part of the QNM into two analogous
frequencies.  
To derive this split, start from a single ray, along which the phase of the 
wave must be constant.
Also suppose that the ray originates from $\theta_-$ and ends at $\theta_+$ 
after traveling only one-half of a cycle of motion in the $\theta$ direction.  
During this time, the statement that the phase is unchanged is that
\begin{equation}
\label{zerophase}
0= - \omega_R T_{\theta}/2   +  (L-|m|)\pi  + m \Delta\phi /2 \,.
\end{equation}
Using (half of) Eq.~\eqref{phiprec}, the real part of the frequency is 
\begin{equation}
\label{totalomega}
\omega_R = L\Omega_\theta(m/L) + m\Omega_{\rm prec}(m/L)\,.
\end{equation}
Note that $\Omega_\theta$ and $\Omega_{\rm prec}$ both depend on $m/L$.

More explicitly, given the orbital parameters $(\mathcal{E},\mathcal{Q},L_z)$,
the quantities $T_\theta$ and $\Delta \phi$ can be obtained by computing
\begin{subequations}\label{eqfreqs}
\begin{align}
T_\theta &=\frac{\partial}{\partial \mathcal E}\oint \sqrt{\Theta} d\theta +\frac{1}{2\Delta}\frac{\partial \mathcal{R}}{\partial \mathcal{E}}\oint \frac{d\theta}{\sqrt{\Theta}}\,, \\
\Delta \phi &=-\frac{1}{L_z}\left[1-\frac{\partial}{\partial\log\mathcal{E}}\right] \oint \sqrt{\Theta} d\theta +\frac{1}{2\Delta}\frac{\partial \mathcal{R}}{\partial L_z}\oint \frac{d\theta}{\sqrt{\Theta}}\,,
\end{align}
\end{subequations}
(expressions that hold for any spherical photon orbit---not simply orbits that 
satisfy the Bohr-Sommerfeld condition) and the two frequencies are given by 
\begin{subequations}
\begin{align}
\Omega_\theta& = 2\pi
\left(\frac{\partial}{\partial \mathcal E}\oint \sqrt{\Theta} d\theta +\frac{1}{2\Delta}\frac{\partial \mathcal{R}}{\partial \mathcal{E}}\oint \frac{d\theta}{\sqrt{\Theta}}\right)^{-1}\\
\Omega_{\rm prec} &= \Omega_\theta\frac{\Delta \phi}{2\pi} -({ \mathrm{sgn} L_z}) \Omega_\theta \,.
\end{align}
 \end{subequations}
These can be expressed in terms of $(\mathcal{E},\mathcal{Q},L_z)$ using 
elliptic integrals (as was done in \cite{Edward}), but we will not carry this 
out explicitly.

For very slowly spinning black holes, a short calculation shows that 
\begin{subequations}
\begin{align}
\Omega_\theta \approx & \frac{1}{\sqrt{27}M} = \sqrt{\frac{M}{r^3_0}}\,,
\label{eq:Omega1} \\
\Omega_{\rm prec} \approx & \frac{2 a}{27 M^2} = \frac{2 S}{r^3_0} \,,
\label{eq:Omega2}
\end{align}
\end{subequations}
where $r_0$ is the circular-photon-orbit radius for a Schwarzschild black hole,
$r_0=3M$, and $S = a M$.
The expression for $\Omega_\theta$ is the Keplerian frequency of the spherical
photon orbit, and $\Omega_{\rm prec} = 2S/r^3_0$ is the Lense-Thirring 
precessional frequency.
In the slow-rotation limit, therefore, our formula recovers Ferrari and 
Mashhoon's result \cite{Ferrari1984}.

\begin{figure}
\includegraphics[width=0.95\columnwidth]{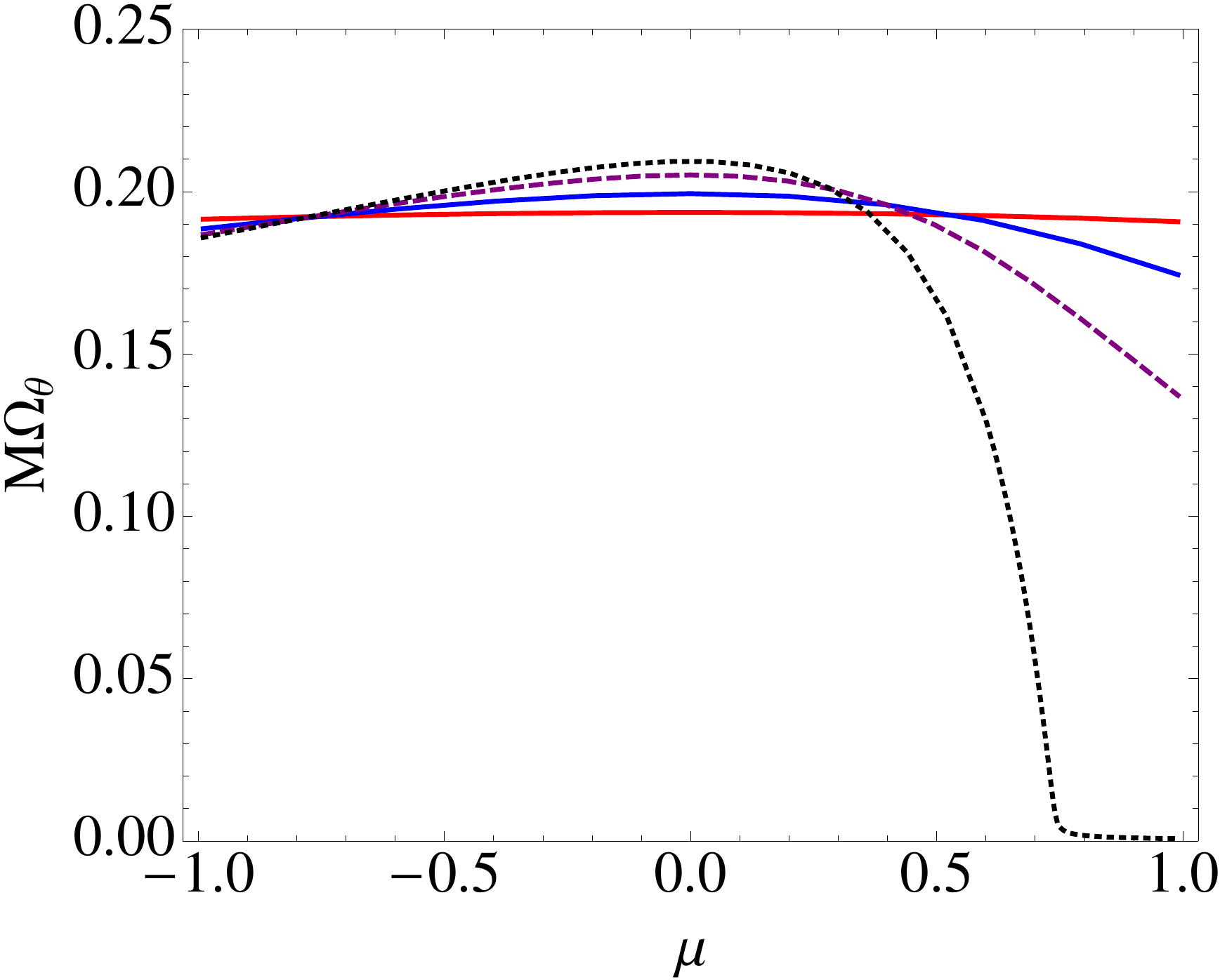}
\caption{Orbital frequency, $\Omega_\theta$, plotted against $\mu$, for 
$a/M=0.3$ [red (light gray) solid curve], $0.7$ [blue (dark gray) solid curve],
0.9 (purple dashed line), and 1 (black dotted line).
The orbital frequency vanishes for a significant range of $\mu$ for extremal
black holes.}
\label{fig:omegatheta}
\end{figure}

\begin{figure}
\includegraphics[width=0.95\columnwidth]{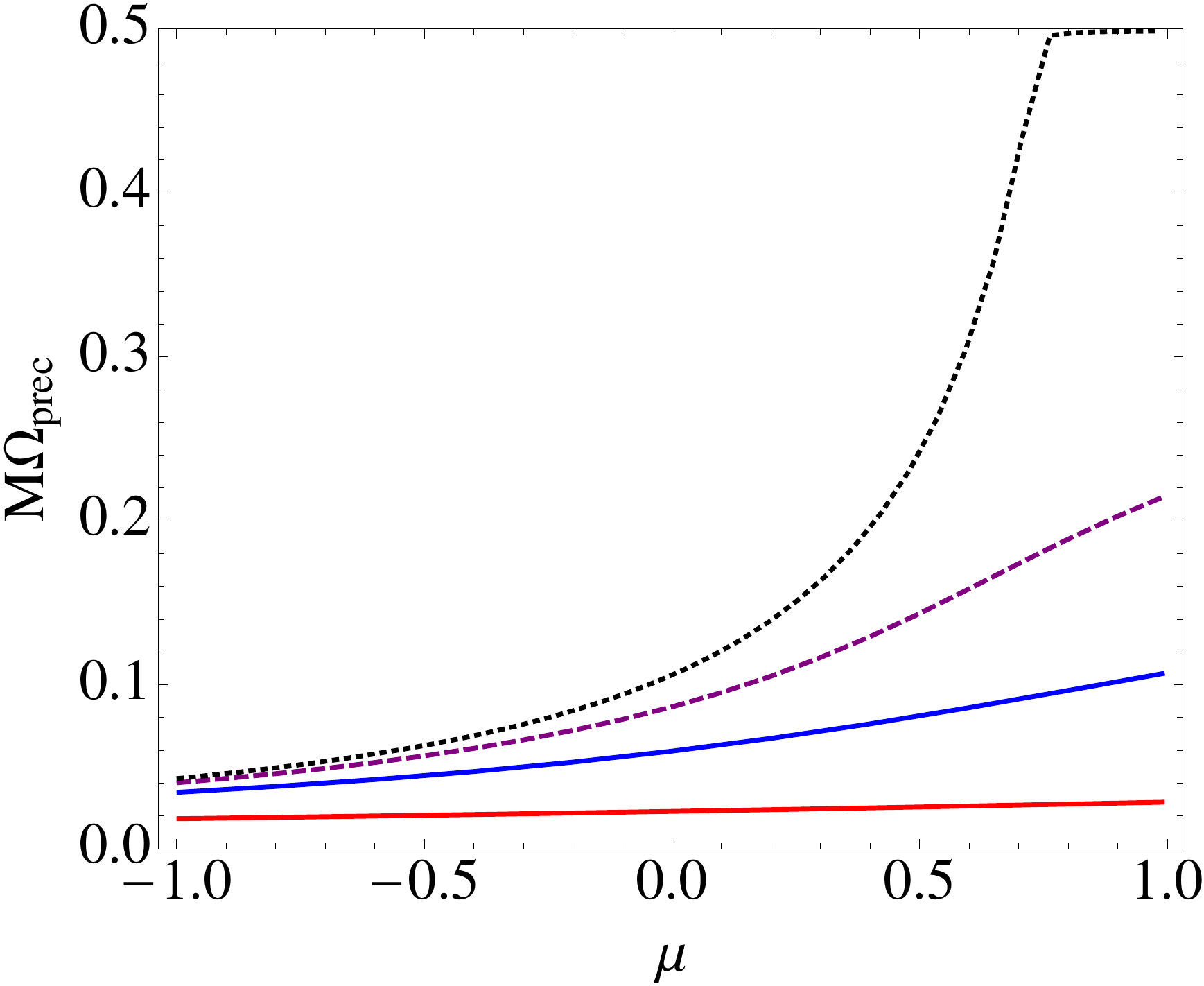}
\caption{Precessional frequency, $\Omega_\phi$, versus $\mu$ plotted 
identically to those curves in Fig.\ \ref{fig:omegatheta} representing the
same black-hole spins.
The precessional frequency approaches the horizon frequency, $\Omega_H$, for
a range of values of $\mu$ for extremal black holes.}
\label{fig:omegaphi}
\end{figure}

For any value of $a$, we can normalize Eq.~\eqref{totalomega} by $L$, and write
\begin{equation}
\Omega_R(a,\mu) =\Omega_\theta(a,\mu) + \mu \Omega_{\rm prec} (a,\mu) \, .
\end{equation}
In Figs.~\ref{fig:omegatheta} and \ref{fig:omegaphi}, we explore the two 
frequencies in the decomposition of $\Omega_R$ by separately plotting 
$\Omega_\theta$ and $\Omega_{\rm prec}$, for different values of $a$.

For small values of $a/M$, $\Omega_\theta$ and $\Omega_{\rm prec}$ are 
consistent with the constant values predicted by Eqs.~\eqref{eq:Omega1} and 
\eqref{eq:Omega2}. 
For larger values of $a/M$, $\Omega_\theta$ does not vary much as a function 
of $\mu$ until $a \sim 0.7 M$; for spins greater than this value, it is only
for larger values of $\mu$ that $\Omega_\theta$ changes significantly by 
decreasing from the equivalent values for $a=0$.
Finally, as $a\rightarrow M$, $\Omega_\theta$ vanishes for 
$\mu \ge \mu_* \approx 0.744$.  
The precessional frequency, $\Omega_{\rm prec}$, on the other hand, 
monotonically increases as a function of $\mu$ even for small values of $a/M$; 
as $a\rightarrow M$, $\Omega_{\rm prec}$ grows to $\Omega_H$ at 
$\mu\sim \mu_*$, and stays there for all values of $\mu > \mu_*$. 
For $a\sim M$ and $\mu >\mu_*$, there is one additional feature worth noting: 
because $\Omega_\theta \sim 0$ and $\Omega_\phi \sim \Omega_H$, this gives rise 
to the interpretation of the mode as a ray that sticks on the horizon and 
corotates with the horizon at its angular frequency of $\Omega_H$; moreover, 
there seems to be no restoring force along the $\theta$ direction.

\subsection{Degenerate quasinormal modes and closed spherical photon orbits}

\begin{figure}
\includegraphics[width=\columnwidth]{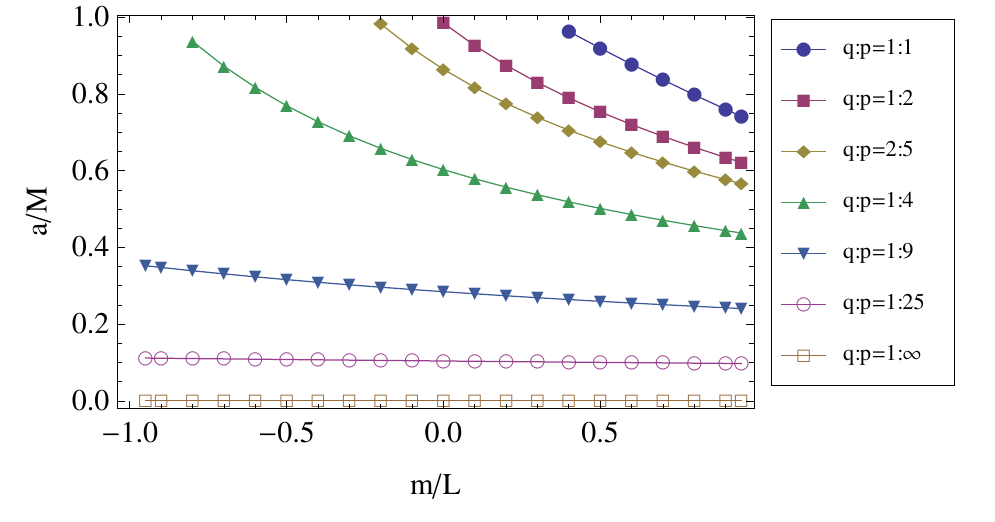}
\caption{A diagram showing the spin parameters, $a$, and the ratios of the 
multipolar indexes $m/L$, at which the orbital and precessional frequencies
have a ratio of $p/q$.
Although we only perform our numerical calculations at a discrete set of $m/L$ 
values (shown by the dots), in the eikonal limit, each set of points for a 
given ratio of $p/q$ approaches a continuous curve.}
\label{fig:PhaseDiagram}
\end{figure}

\begin{figure*}
\includegraphics[width=0.99\textwidth]{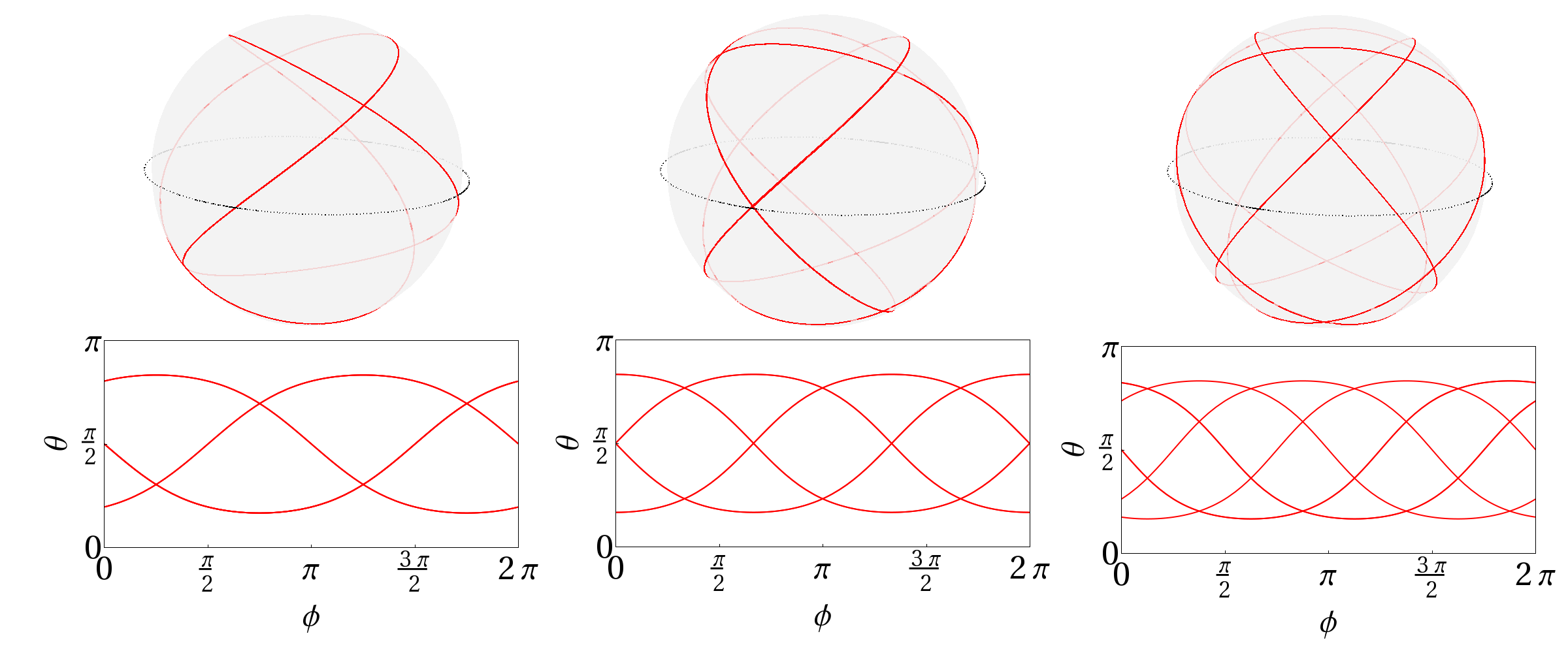}
\caption{For black holes with spins $a/M=0.768$,  $0.612$, and $0.502$, the
spherical photon orbits with $\omega_{\rm orb} = 2\omega_{\rm prec}$, on the 
left, $\omega_{\rm orb} = 3\omega_{\rm prec}$ in the center, and 
$\omega_{\rm orb} = 4\omega_{\rm prec}$ on the right, respectively.
These orbits correspond to quasinormal modes in the eikonal limit with 
$m/L=0.5$.
The top figures show the photon orbit, the red, solid curve, on its photon
sphere (represented by a transparent sphere).
The dashed black line is the equatorial ($\theta=\pi/2$) plane, which was
inserted for reference.
The bottom figures are the same photon orbits, but plotted in the 
$\phi$-$\theta$ plane, instead.}
\label{fig:orbits}
\end{figure*}

Finally, in this section, we interpret the degeneracy of QNM frequencies (of 
which Fig.\ \ref{numdeg} was an example).
Recall that in that figure, for $a/M \approx 0.7$, we found pairs of modes
such as $(2,2)$ and $(3,-2)$, $(3,2)$ and $(4,-2)$, $(4,2)$ and $(5,-2)$, and
so on, all have approximately the same frequency.  
For another, lower spin $a/M \approx 0.4$, pairs like $(3,3)$ and $(4,-3)$, 
$(4,3)$ and $(5,-3)$, et cetera, have approximately the same frequency.

The approximate degeneracy exists because the ratio between $\Omega_\theta$ and
$\Omega_{\rm prec}$ can be rational, and the photon orbits close.  
If for a certain mode of a black hole with spin $a$, with $m$ and $L$, and for
integers $p$ and $q$,
\begin{equation} 
\label{eq:RationalRatio}
q \Omega_\theta\left(a,\frac{m}{L}\right) = 
p \Omega_{\rm prec}\left(a,\frac{m}{L}\right) \,,
\end{equation}
this means that there exists a closed spherical photon orbit that satisfies the
conditions necessary to correspond to a QNM. 
Equation~\eqref{eq:RationalRatio} implies that
\begin{align}
&L \Omega_\theta\left(a, \frac{m}{L}\right)  + m 
\Omega_{\rm prec}\left(a, \frac{m}{L}\right)\nonumber\\
=&
(L+kq) \Omega_\theta\left(a, \frac{m}{L}\right)  + 
(m-kp)\Omega_{\rm prec}\left(a, \frac{m}{L}\right)\,.
\end{align}
If $\Omega_\theta$ and $\Omega_{\rm prec}$ do not change much from $\mu =m/L$ 
to $\mu' = (m-kp)/(L+kq)$ (either because spin is small---and therefore 
$\Omega_\theta$ and $\Omega_{\rm prec}$ depend weakly on $\mu$---or because 
$L \gg kq$ and $m \gg kp$), then
\begin{equation}
\omega_R^{l,m} \approx \omega_R^{l+kq,m-kp} \,.
\end{equation} 
Because $\Omega_I$ depends similarly on $\mu$, under the same conditions,
\begin{equation}
\omega_I^{l,m} \approx \omega_I^{l+kq,m-kp} \, ;
\end{equation}
therefore, the modes are degenerate. 
It is also clear from Eq.~\eqref{eq:RationalRatio} that the degeneracy happens 
at the same time that the corresponding orbit is closed. 
The three series mentioned at the beginning of the paper correspond to 
$p/q =4$, $6$, and $8$, respectively (for $k=1$).

\subsubsection{Slowly spinning black holes}

For $a/M \ll 1$, when Eqs.~\eqref{eq:Omega1} and \eqref{eq:Omega2} apply, the 
condition for degenerate modes becomes
\begin{equation}
\label{reslowa}
\frac{q \sqrt{27}}{2p} = \frac a M \ll 1
\end{equation}
(a statement that holds independent of $\mu$).  
This implies that orbits of all inclinations close for these spins. 

For these specific spins, the QNM spectrum is completely degenerate, by which 
we mean that all modes have the same decay rate, and all real parts of the 
frequencies are integer multiples of only one frequency (similar to those of a 
Schwarzschild black hole).   
Using this approximate formula to find $a$ for the three instances of 
degeneracy in Fig.\ \ref{numdeg}, we find
\begin{equation}
a_{4/1} \approx 0.65 M,\quad
a_{6/1} \approx 0.43 M,\quad
a_{8/1} \approx 0.32 M.
\end{equation}
These are not very far away from spins we found empirically.

\subsubsection{Generic black holes}

For a generic spin parameter $a$, we will explain degeneracies that exist 
around a mode with $L \gg 1$ and $|m| \gg 1$.
If the condition in Eq.~\eqref{eq:RationalRatio} holds for 
$p, q \ll \min(L,|m|)$, then there is a range of 
$|k| \ll \min(L,|m|)/\max(p,q)$ in which there is a degeneracy between all 
$(L+kq,m-kp)$ and $(L,m)$.
These modes must be those close to the mode of indices $(L,m)$, because,
strictly speaking, it is only the orbit corresponding to $m/L$ which is
precisely closed. 

To find this degeneracy, we will search for spin parameters $a$ for which Eq.\
\eqref{eq:RationalRatio} holds for any set of indexes $(L,m)$ and integers 
$(p,q)$ that satisfy $L,|m| \gg p,q$ (we generally either find one or zero 
solutions).
To visualize this degeneracy, for each pair $(p,q)$, we will mark all possible
pairs of $(m/L,a)$ in a 2D plot; the values of the spins are sufficiently dense
for each value of $m/L$ that they form a smooth curve when plotted 
against $m/L$. 
Some of these curves are shown in Fig.~\ref{fig:PhaseDiagram}.  
Because for a fixed $p/q$ the degenerate spins for $a/M \lesssim 0.3$ are 
nearly independent of $m/L$, Eq.~\eqref{reslowa} should be an accurate 
prediction for spins less that that value.
As a concrete illustration of the orbits corresponding to these degenerate 
modes, we plot closed orbits for $m/L=0.5$ and for $a/M \approx 0.5$, $0.61$, 
and $0.77$ in Fig.~\ref{fig:orbits}.
The values of the spins agree quite well with those predicted in Fig.\ 
\ref{fig:PhaseDiagram}.

\section{Conclusions and Discussion}
\label{sec:Conclusions}

In this paper, we extended the results of several earlier works 
\cite{Ferrari1984,Sam,Kokkotas,Iyer} to compute the quasinormal-mode 
frequencies and wave functions of a Kerr black hole of arbitrary astrophysical 
spins, in the eikonal limit ($l\gg 1$).
We focused on developing a greater intuitive understanding of their behavior,
but, in the process, we calculated expressions for large-$l$ quasinormal-mode 
frequencies that are reasonably accurate even at low $l$.
Specifically, we applied a WKB analysis to the system of equations defined by 
the angular and radial Teukolsky equations.
Using a Bohr-Sommerfeld condition for the angular equation, we related the 
angular separation constant to the frequency; when we expanded the constraint
to leading order in $a\omega/l$, we found an equally accurate and algebraically
simpler relation between the frequency and angular eigenvalue.
We then used a well-known WKB analysis on the radial Teukolsky equation to 
obtain expressions for the QNM frequencies and the angular separation 
constants. 
The accuracy of the approximate expressions for the QNM frequency are observed 
to be of order $O(L^{-2})$ even though we had only expected a $O(L^{-1})$ 
convergence for the imaginary part. 

Next, we reviewed that a massless scalar wave in the leading-order, 
geometric-optics approximation obeys the Hamilton-Jacobi equations, which are 
very similar to the Teukolsky equations when $l\gg1$.
By identifying terms in the Hamilton-Jacobi equations and Teukolsky equations,
we related the conserved quantities of the Hamilton-Jacobi equations to the
eigenvalues of the separated Teukolsky equations.
Specifically, we confirmed that the energy, angular momentum in the $z$ 
direction, and Carter constant in the Hamilton-Jacobi equations correspond
to the real frequency, the index $m$, and the angular eigenvalue minus $m^2$
in the Teukolsky equations, respectively.
Furthermore, we found that the conditions that define a quasinormal mode in the
WKB approximation are equivalent to the conditions in the geometric-optics
approximation that determine a spherical photon orbit that satisfies an 
identical Bohr-Sommerfeld quantization condition.

By analyzing the next-to-leading-order, geometric-optics approximation, we 
showed that the corrections to the amplitude of the scalar wave correspond to
the imaginary parts of the WKB quantities.
Specifically, we saw that the imaginary part of the frequency is equal to a
positive half-integer times the Lyapunov exponent averaged over a period of
motion in the $\theta$ direction.
The imaginary part of the angular eigenvalue is equal to the imaginary part of
the Carter constant, which is, in turn, related to an amplitude correction to
geometric-optics approximation to the angular function for $\theta$.

We then applied these results to study properties of the QNM spectra of Kerr 
black holes.
We observed that for extremal Kerr black holes a significant fraction of the 
QNMs have nearly zero imaginary part (vanishing damping) and their 
corresponding spherical photon orbits are stuck on the horizon (in 
Boyer-Lindquist coordinates).
We plan to study this unusual feature of extremal Kerr black holes in future 
work.
In addition, we showed that for Kerr black holes of any spin, the modes' 
frequencies (in the eikonal limit) are a linear combination of the orbital and 
precession frequencies of the corresponding spherical photon orbits. 
This allows us to study an intriguing feature of the QNM spectrum: namely, 
when the orbital and precession frequencies are rationally related---i.e, when 
the spherical photon orbits are closed---then the corresponding 
quasinormal-mode frequencies are also degenerate.

We hope that the approximate expressions for the quasinormal-mode frequencies
in this paper will prove helpful for understanding wave propagation in the Kerr
spacetime.
This not unreasonable to suppose, because Dolan and Ottewill have shown in 
\cite{Sam2,Sam3} that to calculate the Green's function analytically in the 
Schwarzschild spacetime, one needs to know analytical expressions for the 
frequency of the quasinormal modes (specifically, this comes from the fact that
the frequencies of the quasinormal mode are the poles of the Green's function 
in the frequency domain).
We, therefore, think that our approximate formulas could assist with the
calculation of the Green's function in the Kerr spacetime, in future work.

\begin{acknowledgments}
We thank Emanuele Berti for discussing this work with us and pointing out 
several references to us.
We also thank Jeandrew Brink for insightful discussions about spherical photon 
orbits in the Kerr spacetime. 
We base our numerical calculation of the QNM frequencies on the Mathematica 
notebook provided by Emanuele Berti and Vitor Cardoso \cite{bert3}.
This research is funded by NSF Grants PHY-1068881, PHY-1005655, 
CAREER Grant PHY-0956189; NASA Grant No.NNX09AF97G; the Sherman Fairchild 
Foundation, the Brinson Foundation, and the David and Barabara Groce Startup 
Fund at Caltech.
\end{acknowledgments}

\appendix
\section{The Taylor expanded Bohr-Sommerfeld condition} 
\label{sec:BohrSommerfeldAp}

The Bohr-Sommerfeld constraint \eqref{Aeq} gives us a way to evaluate $A_{lm}$ 
in terms of $l$, $m$, and $\omega$; the error in this approximation scales as 
$1/l$. 
Because it is an integral equation, it is not particularly convenient to solve,
and it is beneficial to have an approximate, but algebraic expression for the 
frequency of a QNM.
With the benefit of hindsight, one can confirm through numerical calculations
of exact QNM frequencies performed using Leaver's method that the parameter
$a\omega/l$ is numerically a small number for all black hole spins.
We can then expand the angular separation constant, $A_{lm}$, in a series in
$a\omega/l$ as $A_{lm}=A^0_{lm}+\delta A_{lm}$, where $A^0_{lm}$ satisfies the 
equation
\begin{equation}
\int^{\theta^0_{+}}_{\theta^0_{-}}\sqrt{A^0_{lm}-\frac{m^2}{\sin^2\theta}}
=\pi\left (l+\frac{1}{2}-|m| \right ) \, ,
\label{eq:ZeroOrderAlm}
\end{equation} 
and at leading order, $\theta^0_{+},\theta^0_{-}=\pm \arcsin[ m/(l+1/2)]$. 
One can easily verify that the solution to this equation is the angular 
eigenvalue of a Schwarzschild black hole, $A^0_{lm}=(l+1/2)^2$ (note that we 
are assuming $l \gg 1$). 
Now we will compute the lowest-order perturbation in $a\omega/l$, which turns
out to be quadratic in this parameter [i.e., $(a\omega/l)^2$] below:
\begin{equation}
\label{eqperta}
\int^{\theta^0_{+}}_{\theta^0_{-}}\frac{\delta A_{lm}+a^2\omega^2\cos^2\theta}
{\sqrt{(l+1/2)^2-m^2/\sin^2\theta}}d\theta=0 \, .
\end{equation}

The integration limits $\theta_{+}, \theta_{-}$ also can be expanded in a
series in $a\omega/l$, and the lowest-order terms of this series are given by 
$\theta^0_{+},\theta^0_{-}$; 
The perturbation in $\theta_{+}, \theta_{-}$ would result in some quartic 
corrections in $a\omega/l$ [i.e., $(a\omega/l)^4$] when we evaluate the 
integrals of Eqs.\ (\ref{eqperta}) and (\ref{eq:ZeroOrderAlm}), because the
integrand is of order $(a\omega/l)^2$ and the width of the correction in 
$\theta_{+}, \theta_{-}$ are also of order $(a\omega/l)^2$.
As a result, we will not need it here. 
Evaluating the integral in Eq.\ (\ref{eqperta}) is straightforward, and we find
\begin{equation}
\label{qntheta2}
A_{lm}=A^0_{lm}+\delta A_{lm} =l(l+1)-\frac{a^2\omega^2}{2}
\left [1-\frac{m^2}{l(l+1)}\right ]
\end{equation}

Interestingly, the above expression is consistent with the expansion of 
$A_{lm}$ for small $a\omega$ given in \cite{Berticasal}, even in the eikonal 
limit, where $a\omega$ is large.
The reason for this fortuitous agreement is again that for QNMs of Kerr black 
holes of any spin, $a\omega/l$ is small, and the expansion only involves even 
powers of this parameter, $(a\omega/l)^2$.
Because the coefficients in the expansion of $a\omega$ scale as $1/l^k$ for
even powers of $(a\omega)^k$ and $1/l^{k+1}$ for odd powers of $(a\omega)^k$,
in the limit of large $l$, the two series actually are equivalent in the
eikonal limit.
In principle, one can also expand and solve Eq.~\eqref{Aeq} to higher 
orders in the parameter $a\omega/l$ and compare with the expansion in $a\omega$
in \cite{Berticasal}; we expect that the two series should agree.
This would be useful, because it would effectively let one use the small 
$a\omega$ expansion in the eikonal limit, where the series would, ostensibly,
not be valid.

\end{document}